\documentclass[reprint,aps,prl,amsmath,amssymb,longbibliography, notitlepage]{revtex4-2}
\usepackage{amsmath,amssymb,bm,graphicx}
\usepackage{graphics}
\usepackage{float} 
\usepackage{algorithm}
\usepackage[stable]{footmisc}
\usepackage{soul}
\usepackage{physics}
\usepackage{pifont}
\usepackage[normalem]{ulem}
\usepackage[colorlinks, linkcolor= blue, citecolor = blue, urlcolor=blue]{hyperref}

\newcommand{\cmark}{\ding{51}}%
\newcommand{\xmark}{\ding{55}}%
\def\be{\begin{equation}}
\def\ee{\end{equation}}
\def \bea{\begin{eqnarray}}
\def \eea{\end{eqnarray}}
\def \nn{\nonumber}

\begin{document}
\title{Planar Hall Effect in Quasi-Two-Dimensional Materials}

\author{Koushik Ghorai}
\thanks{KG and SD contributed equally and are joint first authors.}
\author{Sunit Das}
\thanks{KG and SD contributed equally and are joint first authors.}
\author{Harsh Varshney}
\author{Amit Agarwal}
\email{amitag@iitk.ac.in}
\affiliation{Department of Physics, Indian Institute of Technology Kanpur, Kanpur-208016, India}

\begin{abstract}
The planar Hall effect in 3D systems is an effective probe for their Berry curvature, topology, and electronic properties. However, the Berry curvature-induced conventional planar Hall effect is forbidden in 2D systems as the out-of-plane Berry curvature cannot couple to the band velocity of the electrons moving in the 2D plane. Here, we demonstrate a unique 2D planar Hall effect (2DPHE) originating from the hidden planar components of the Berry curvature and orbital magnetic moment in quasi-2D materials. We identify all planar band geometric contributions to 2DPHE and classify their crystalline symmetry restrictions. Using gated bilayer graphene as an example, we show that in addition to capturing the hidden band geometric effects, 2DPHE is also sensitive to the Lifshitz transitions. Our work motivates further exploration of hidden planar band geometry-induced 2DPHE and related transport phenomena for innovative applications.
\end{abstract}

\maketitle


The planar Hall effect (PHE) is the generation of longitudinal and transverse voltages in the plane of the applied electric  ($\bm E$) and magnetic fields ($\bm B$). In contrast to the conventional and anomalous Hall effect, the transport in PHE is dissipative, and the response typically varies quadratically with the $B$. PHE has extensive applications in magnetic sensors and memory devices~\cite{Quynh_19}. 
In 3D materials, PHE generally originates from the 
coupling of the Berry curvature (BC) and orbital magnetic moment (OMM) to the band velocity and in-plane magnetic field, which generates a longitudinal and transverse planar response. Initial studies of PHE used it effectively to probe the magnetization reversal in magnetic materials~\cite{Yau_jp71, Li_jp10, Roukes_prl03, Ploong_prb05, Schaefer_prb21}. More recently, we have used PHE to explore novel topological semimetals ~\cite{Nandy_prl17,shekhar_prb18,deng_prb19, Yao_prb23, Xie_prb19, Kamal_prr20_chiral, Weng_prb23, Kamal_prb19, Sunit_prb23, Ipsita_24, Ipsita_physica24} and topological insulators~\cite{ Chang_prb18, Nandy_sr18}. 

However, conventional PHE probes are ineffective in 2D  systems. As the 2D plane confines the orbital motion of electrons, these systems can host only out-of-plane Berry curvature and orbital magnetic moment~\cite{Xiao_rmp10, Cayssol_JPM21}. Consequently, the PHE induced by the component of the BC and OMM along the applied in-plane magnetic field is forbidden in perfect 2D systems. Some 2D materials with strong spin-orbit coupling exhibit an intrinsic magneto-Hall response driven by magnetic field-induced changes to the Berry curvature~\cite{Zyuzin_prb20, Culcer_prl21, Carmine_prrL21, Liang_NP18, Zhou_nature22, Dai_prbL22, Shengyuan22_in-plane}. However, such responses are antisymmetric tensors and absent in systems lacking strong spin-orbit interactions. These limitations severely restrict our ability to utilize PHE to explore fundamental physics and develop ultra-sensitive magnetic sensors and other applications in 2D materials. 

\begin{figure}
    \centering
    \includegraphics[width=\linewidth]{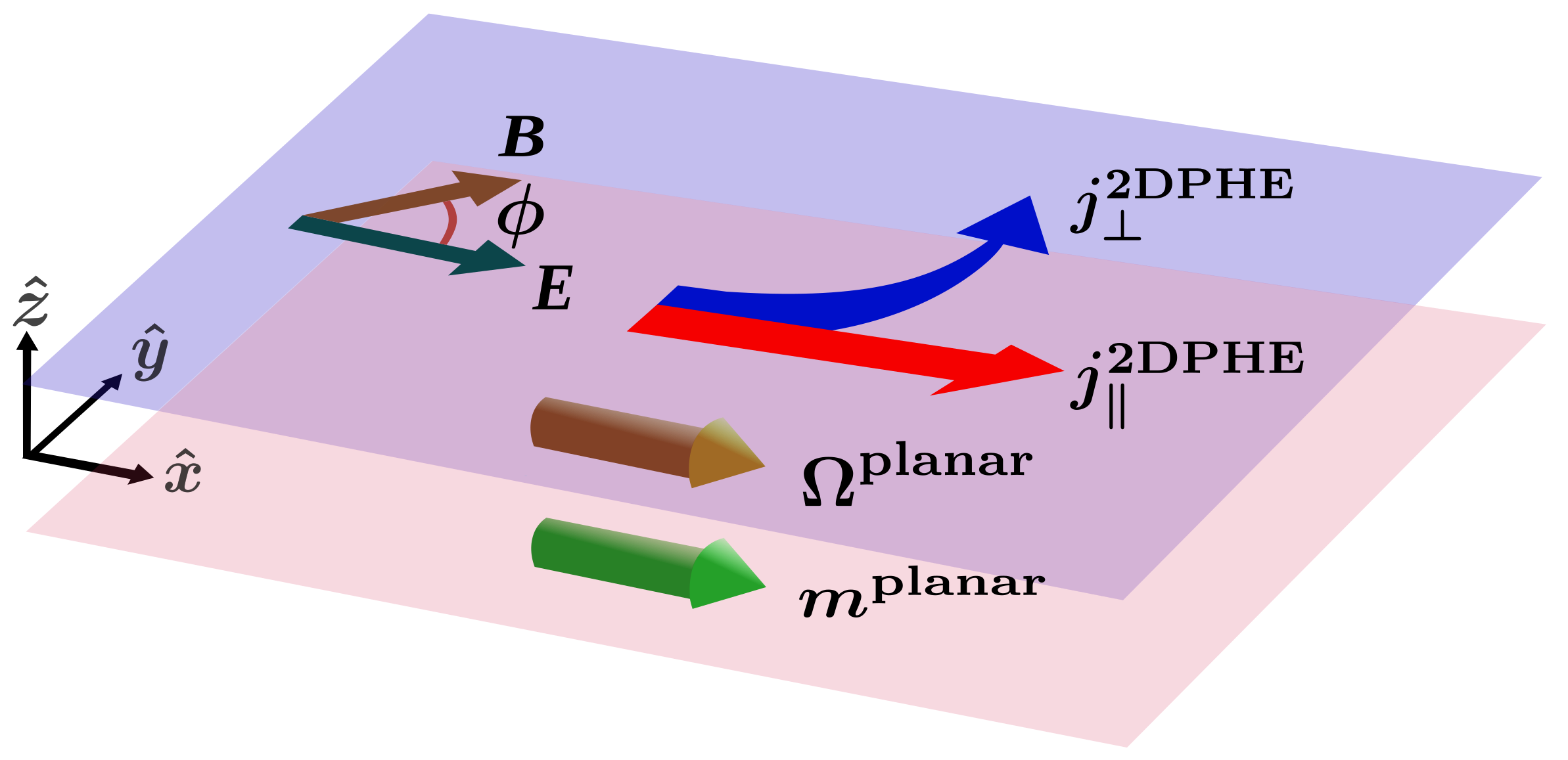}
    \caption{Schematic for 2D planar Hall effect (2DPHE). Layered 2D materials host hidden planar Berry curvature (${\bm \Omega}^{\rm planar}$) and planar orbital magnetic moment (${\bm m}^{\rm planar}$) arising from inter-layer tunneling. The ${\bm \Omega}^{\rm planar}$ and ${\bm m}^{\rm planar}$ combine with the in-plane electric and magnetic field to induce a longitudinal and transverse current in the 2D plane. \label{Fig_1}}
\end{figure}

In this Letter, we introduce a unique 2D planar Hall effect (2DPHE) in layered 2D materials such as bilayer graphene. 
Layered 2D materials with finite inter-layer tunneling can host an intrinsic in-plane component of the BC and OMM if the system's space inversion or time-reversal symmetry is broken~\cite{Resta_prb20,murakami_prb20, Jin_prbL21}. We demonstrate that these relatively unexplored in-plane components of the band geometric quantities induce the 2DPHE response (see Fig.~\ref{Fig_1}). We present a thorough analysis of the 2DPHE responses, particularly their angular variation (angle between ${\bm E}$ and ${\bm B}$), and classify the crystalline symmetry restrictions on the different 2DPHE response tensors.  
As an illustrative example, we focus on Bernal stacked bilayer graphene to demonstrate a sizable and gate-tunable 2DPHE  response. Beyond predicting the unique phenomena of 2DPHE, our findings motivate the exploration of other transport and optical phenomena induced by the hidden planar band geometric quantities in layered 2D materials~\cite{Duong_ACS17_vdW}. 

\begingroup
\setlength{\tabcolsep}{6.5pt}
\begin{table*}[t!]
\caption{ The symmetry restrictions of the longitudinal and planar Hall response tensors pertaining to two-dimensional materials. The cross (\xmark) and the tick (\cmark) mark signify the corresponding response tensor is symmetry forbidden and allowed, respectively. The longitudinal and transverse PHE tensors in the same row have identical symmetry restrictions.}
\begin{tabular}{c c c c c c c c c c c c c c c c}
\hline \hline 
\noalign{\vskip 2pt}
\rm{Longitudinal } & \rm{Transverse } & ${\mathcal P}$ & ${\mathcal T}$ & $\mathcal{P}\mathcal{T}$  &  ${\cal M}_x$ & ${\cal M}_y$ & ${\cal M}_z$ & ${\cal C}_{2x}$ & ${\cal C}_{2y}$ & ${\cal C}_{2z}$ & ${\cal C}_{3z}$ & ${\cal C}_{4z}$ & ${\cal C}_{6z}$ & ${\cal S}_{4z}$  & ${\cal S}_{6z}$ \\
\noalign{\vskip 2pt}
\hline \hline 

$\chi_{xx;x}$ & $\chi_{yx;y}$ & \cmark  & \xmark  & \xmark & \cmark &  \xmark & \xmark & \cmark & \xmark & \xmark &  \cmark  & \xmark & \xmark & \xmark & \cmark \\

$\chi_{xx;y}$ & $\chi_{yx;x}$ & \cmark  & \xmark & \xmark  & \xmark &  \cmark & \xmark & \xmark & \cmark  & \xmark & \cmark  & \xmark & \xmark & \xmark & \cmark \\

\noalign{\vskip 2pt}
\hline
\noalign{\vskip 1pt}

$\chi_{xx;xx}$, $\chi_{xx;yy}$ & $\chi_{yx;xy}$ &  \cmark & \cmark & \cmark & \cmark & \cmark & \cmark &  \cmark &  \cmark &  \cmark &  \cmark &  \cmark &  \cmark &  \cmark &  \cmark \\ 

$\chi_{xx;xy}$ & $\chi_{yx;xx}$, $\chi_{yx;yy}$ &  \cmark & \cmark & \cmark & \xmark & \xmark & \cmark & \xmark & \xmark & \cmark & \cmark & \cmark & \cmark & \cmark & \cmark \\

\noalign{\vskip 2pt}
\hline \hline
\end{tabular}
\label{table_1}
\end{table*}
\endgroup

{\it Planar-BC and OMM in 2D systems:--}
In quasi-2D materials with two or more atomic layers, the finite inter-layer hopping amplitude enables the inter-layer tunneling of electrons. The inter-layer coherence of the electron wavefunction gives rise to hidden planar components of the BC and OMM. These are given by, 
\bea \label{2D_BC}
{\bm \Omega}_{n\bm k}^{\rm planar} &&= {2}{\hbar} ~{\rm Re} \sum_{n' \neq n} \frac{{\bm v}_{nn'} \times \bm{\mathcal{Z}}_{n' n}}{(\varepsilon_{n \bm k} -\varepsilon_{n' \bm k})}~, \\
{\bm m}_{n\bm k}^{\rm planar} &&=  {e}~{\rm Re} \sum_{n' \neq n} {\bm v}_{n n'} \times \bm{\mathcal{ Z}}_{n' n}~. \label{2D_OMM}
\eea
Here, we have defined the velocity matrix elements as $\hbar{\bm v}_{nn'} = \bra{u_{n{\bm k}}} \nabla_{\bm k}  \mathcal{H} \ket{u_{n'{\bm k}}}$, with ${\bm k}=(k_x,k_y)$.  $\varepsilon_{n \bm k}$ and $\ket{u_{n{\bm k}}}$ are the band energy and periodic part of the Bloch wavefunction for the $n$-th band, respectively. The matrix elements of the out-of-plane position operator are defined as $\bm{\mathcal{Z}}_{nn'}=\hat{\bm z} 
 \bra{u_{n\bm k}} \mathcal{Z} \ket{u_{n' \bm k}}$, with $\mathcal{Z}$ being the position operator along the $\hat{\bm z}$-direction. These in-plane contributions combined with the conventional out-of-plane components yield the total BC and OMM for quasi-2D materials. We present the detailed derivation of these equations in Sec.~S1 of the Supplementary Materials (SM)~\footnote{The \href{https://www.dropbox.com/scl/fi/fxb532ylx6dk5xicrcgea/SM_2DPHE.pdf?rlkey=r2ajvbgg2j6t5y3z4xbwihw3u&st=9av0zfu2&dl=0}{Supplementary Material} discusses: i) the derivation of planar-BC and planar-OMM expressions, ii) general expression for planar-BC and planar-OMM and analytical calculation of them for $2\times2$ low-energy bilayer graphene model. iii) the detailed derivation of longitudinal and planar Hall response tensors, iv) the details of symmetry analysis, v) the strain implementation in the tight-binding model for bilayer graphene and $y$ components of planar Berry curvature and OMM, vi) the Van Hove singularity and Lifshitz transition of Fermi surface, vii) estimation of planar Hall voltage, viii) comparison of planar Berry curvature and OMM for bilayer and trilayer graphene, and ix) other in-plane magneto-Hall responses in two-dimensional systems.}. We emphasize that the planar-BC and planar-OMM rely on inter-layer hybridization of electronic states, which makes the off-diagonal components of $\bm{\mathcal{Z}}_{nn'}$ finite. We illustrate the emergence of the planar-BC and OMM and their symmetry properties in an intuitive way using a $2\times2$ low energy model~\cite{Schaefer_prb21, Chen_NE21} Hamiltonian of bilayer graphene in Sec.~S2 of the SM~\cite{Note1}.

{\it 2D Planar Hall effect:--}
In 2D systems, generally, the magnetic field interacts with electrons primarily through Zeeman coupling to its spin~\cite{Zyuzin_prb20, Culcer_prl21, Carmine_prrL21, Liang_NP18, Zhou_nature22, Shengyuan22_in-plane}. In contrast, the planar-OMM allows the magnetic field to couple directly to the orbital motion of electrons. This modifies the band energy ($\tilde{\varepsilon}_{n\bm k} = \varepsilon_{n\bm k} -{{\bm m}_{n\bm k}^{\textrm{planar}}}\cdot {\bm B}$) and the band velocity. More importantly, the planar-BC combines with the band velocity to generate a finite chiral magnetic velocity\cite{Das_PRR_2020a} in 2D systems, which is $\propto ({\bm v}_{n\bm k}\cdot{{\bm \Omega}_{n\bm k}^{\textrm{planar}}}) {\bm B}$. We show below that these magnetic field-dependent velocity contributions arising from the hidden planar-BC and -OMM generate a previously unexplored planar Hall effect in 2D systems. 

In the semiclassical Boltzmann transport framework, the charge current is given by ${\bm j} = -e \sum_n \int [d {\bm k}] { \dot{\bm r}_n} g_{n\bm k}$. Here $g_{n\bm k}$ is the non-equilibrium distribution function, $\dot{\bm r}_n$ is the wave-packet velocity, and $[d {\bm k}] \equiv d^2{\bm k}/{(2\pi)^2}$ for 2D systems. Using the expressions of the planar-BC and OMM modified $\dot{\bm r}_n$ and $g_{n\bm k}$ up to linear order in the applied electric field, we calculate 
the planar current density to the first and second orders in the magnetic field strength $B$. See Sec.~S3 of the SM for a detailed derivation~\cite{Note1}. We obtain the longitudinal and transverse components of the 2DPHE currents to be %
\be \label{response_tensors}
j_a = \tau \chi_{ab;c} E_b B_c + \tau \chi_{ab;cd} E_b B_c B_d~.
\ee%
Here, $\tau$ is the electron scattering time, $\{a,b,c,d\}\in \{x,y\}$ are the 2D Cartesian coordinates, and the Einstein summation convention is used. In Eq.~\eqref{response_tensors}, $\chi_{ab;cd}$ is symmetric under the exchange of magnetic field indices $(c,d)$. 
The 2DPHE response tensors can expressed as a sum of the planar-BC, planar-OMM, and mixed terms, 
\be \label{eq_sum}
\chi_{ab;c(d)} = \chi_{ab;c(d)}^{\textrm{BC}} + \chi_{ab;c(d)}^{\textrm{OMM}}+\chi_{ab;c(d)}^{\textrm{BC+OMM}}~.
\ee
We obtain the planar-BC contributions to be
%
%
\bea \label{chi_abc}
    \chi_{ab;c}^{\textrm{BC}} &=&  -e^2 \int_{n,{\bm k}} [(v_a\delta_{bc}+v_b\delta_{ac})\Omega_V - \frac{e}{\hbar}v_a v_b\Omega_c] f_0', \label{chi_B1}  \\
    \label{chi_abcd}
     \chi_{ab;cd}^{\textrm{BC}} &=& -\frac{e^2}{2} \int_{n,{\bm k}} \biggl[ \delta_{ad}\delta_{bc}\Omega_{V}^2 - \frac{e}{\hbar}(v_a\delta_{bc}+v_b\delta_{ac})\Omega_d\Omega_{V} \nn \\
     &&  +  \frac{e^2}{\hbar^2}v_av_b\Omega_c\Omega_d \biggr] f_0' + (c\leftrightarrow d)~. \label{chi_B2}
\eea  
Here, $\Omega_V \equiv (e/\hbar)\bm{v}_{\bm k}\cdot \bm{\Omega}_{\bm k}$ with $\hbar {\bm v}_{\bm k}= \nabla_{\bm k}{ \varepsilon}_{\bm k}$ being the band velocity without any magnetic field, and $\delta_{ab}$ is the Kronecker delta function. For brevity, we have defined $\int_{n, {\bm k}} \equiv \sum_n \int [d{\bm k}]$, and we do not explicitly mention the band index $n$ in the physical quantities. 
We present the expressions for other contributions in Eq.~\eqref{eq_sum} in Sec.~S3 of the SM~\cite{Note1}.

The planar response tensors in Eqs.~\eqref{chi_abc} and ~\eqref{chi_abcd} are proportional to either ${\bm v}_{\bm k}\cdot{\bm \Omega}_{\bm k}$ or, $\hat{\bm B}\cdot{\bm \Omega}_{\bm k}$ or, the combination of these terms. For a quasi-2D system, all of these terms vanish if we consider only the conventional out-of-plane BC. As a consequence, earlier works missed this phenomenon. This highlights the crucial role of the hidden planar-BC and -OMM in generating the PHE response in quasi-2D systems. 
Furthermore, the response tensors $\chi_{ab;c}$ and $\chi_{ab;cd}$ are symmetric with respect to its first two indices. Therefore, we have $\sum_a j_a E_a \neq 0$, indicating the dissipative nature of the planar Hall current. Having established the possibility of 2DPHE, we now analyze the restrictions imposed by crystalline point group symmetries on different 2DPHE response tensors.

{\it Crystal symmetry restrictions:---}
The inversion symmetry ($\cal P$) imposes no constraints as both $\chi_{ab;c}$ and $\chi_{ab;cd}$ represent linear in $E$ responses.
However, under time-reversal ($\cal T$) operation, ${j}$, ${B}$, and $\tau$ reverse signs, while $E$ is $\cal T$ even. Thus, $\chi_{ab;c}$ is a third rank  $\cal T$-odd axial tensor~\cite{newnham_symmetry,Gallego_cryst19}, which is forbidden in non-magnetic systems (see Sec.~S4 of SM~\cite{Note1} for details). In contrast, $\chi_{ab;cd}$ is a fourth rank $\cal T$-even polar tensor and is the leading order contribution in non-magnetic systems. 
Denoting a general point group operation via $\cal O$, the $\chi_{ab;c}$ and $\chi_{ab;cd}$ tensors obey the following transformation rules~\cite{newnham_symmetry}
\bea
\chi_{a'b';c'} &=& \eta_{\mathcal{T}}\det{\mathcal O}{\mathcal O}_{a'a}{\mathcal O}_{b'b}{\mathcal O}_{c'c}~\chi_{ab;c}~, \\
\chi_{a'b';c'd'} &=&  {\mathcal O}_{a'a}{\mathcal O}_{b'b}{\mathcal O}_{c'c}{\mathcal O}_{d'd} ~\chi_{ab;cd}~.
\eea%
Here, $\eta_{\cal T}=\pm 1$ is associated with the magnetic point group symmetry transformation: 
$\eta_{\cal T}=-1$ ($\eta_{\cal T}=1$) for magnetic (non-magnetic) point group operation ${\cal O} \equiv {\cal RT}$ (${\cal O} \equiv {\cal R}$),  with $\cal R$ being a spatial operation.

 To be specific about the symmetry constraints of the 2DPHE response, we apply the $\bm E$ along $\hat{\bm x}$ direction, and an in-plane magnetic field at an angle $\phi$ with ${\bm E}$, {\it i.e.}, $(B_x, B_y)=B(\cos\phi, \sin\phi)$ (see Fig.~\ref{Fig_1}).
 As the response tensors are symmetric in the first two indices, the independent tensor elements for the $B$-linear longitudinal (transverse) responses are $\chi_{xx;x}$ and $\chi_{xx;y}$ ($\chi_{yx;x}$ and $\chi_{yx;y}$). 
 The fourth rank tensor $\chi_{ab;cd}$ is symmetric in the first two $(a,b)$ and the last two $(c,d)$ indices. Hence, for the quadratic-$B$ longitudinal (transverse) response $\chi_{xx;xx}$, $\chi_{xx;yy}$ and $\chi_{xx;xy}$ ($\chi_{yx;xx}$, $\chi_{yx;yy}$ and $\chi_{yx;xy}$) are the only independent elements. 
We present crystalline symmetry imposed restrictions on these tensor elements for non-magnetic and magnetic systems in Table~\ref{table_1} and Table~S1 of SM~\cite{Note1}, respectively. An interesting conclusion from our symmetry analysis is that the presence of ${\cal C}_{3z}$ symmetry does not restrict any of the 2DPHE response tensors. This makes hexagonal systems such as multi-layered graphene, transition metal dichalcogenides, and their twisted moir\'e heterostructures good candidates to observe  2DPHE. 
We now focus on the angular variation of the 2DPHE current. 

%
\begin{figure}[t]
    \centering
    \includegraphics[width=\linewidth]{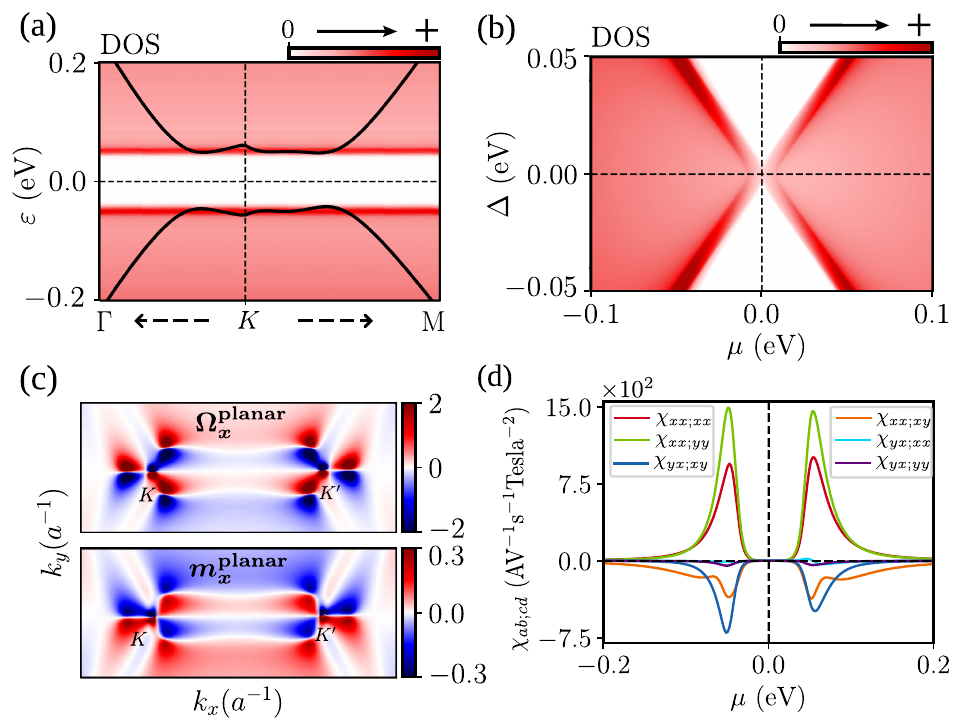}
    \caption{ (a) Electronic band structure of strained BLG around the $K$ point, with the background color showing the density of states (DOS). b) The variation of DOS with chemical potential $\mu$ and the inter-layer potential $\Delta$ in (b). The Van Hove singularity in the DOS near the band edges [also seen in (a)] is accompanied by a Lifshitz transition, where the system evolves from hosting three Fermi pockets around the $K$ (or $K'$) point to one Fermi pocket.
    (c) The upper (lower) panel shows the $k$-space distribution of the $x$-component of the planar BC (OMM) for the first conduction band in the unit of $a^2$ $\left(\frac{e}{\hbar}a^2{\rm eV}\right)$, where $a$ is the lattice constant. (d) Different components of 2DPHE response tensors $\chi_{ab;cd}$ as a function of $\mu$ evaluated at temperature $T = 50$K. 
    }
    \label{Fig_2}
\end{figure}
{\it Angular variation of the 2DPHE currents:--} 
The variation of the 2DPHE currents with the planar angle between the $\bm E$ and $\bm B$ is important for exploring its origin and the relative contribution of different terms. We work with the field configuration described in Fig.~\ref{Fig_1} to obtain the longitudinal and transverse 2DPHE currents: $j_{\parallel(\perp)}^{{\rm 2DPHE}} = \sigma_{{\parallel}(\perp)} E_\parallel $. We calculate the angular dependence of the  2DPHE conductivities to be, 
\bea \label{sigma_long}
\sigma_{\parallel} &=& \tau B(\chi_{xx;x} \cos\phi + \chi_{xx;y}\sin\phi) + \tau B^2 (\chi_{xx;xx} \cos^2{\phi} \nn \\
&& +~\chi_{xx;yy} \sin^2{\phi} + \chi_{xx;xy}\sin{\phi}\cos{\phi})~,
\eea
\bea
\sigma_{\perp} &=& \tau B (\chi_{yx;x} \cos{\phi}+\chi_{yx;y} \sin{\phi}) + \tau B^2 (\chi_{yx;xx} \cos^2{\phi} \nn \\
&& +~\chi_{yx;yy} \sin^2{\phi} + \chi_{yx;xy}\sin{\phi}\cos{\phi})~. \label{sigma_hall}
\eea
These equations and the symmetry restrictions in Table~\ref{table_1} {(and Table S1 in SM~\cite{Note1})} provide a complete characterization of the 2DPHE responses. For non-magnetic systems with an in-plane mirror or an in-plane two-fold rotation symmetry, the longitudinal (transverse) response is entirely captured by $\chi_{xx;xx}$, and $\chi_{xx;yy}$ ($\chi_{yx;xy}$) with the conventional $\cos^2\phi$ ($\sin2\phi$) angular dependence.

{\it 2DPHE in gated bilayer graphene:--}
To demonstrate the 2DPHE in a realistic system, we consider the tight-binding model of Bernal stacked bilayer graphene (BLG). It offers the natural advantage of being readily available, and its doping and layer asymmetry can be tuned via the combination of top and bottom gate voltages. The Hamiltonian of pristine BLG with a vertical displacement field possesses $\cal T$ and ${\cal C}_{3z}$ symmetry, while it breaks $\cal P$ symmetry. Owing to the presence of $\cal T$ symmetry in BLG, only the $B^2$ contributions to the 2DPHE in BLG are allowed. The breakdown of $\cal P$ symmetry is crucial for inducing a planar-BC and planar-OMM in systems with $\cal T$ symmetry. BLG also has a mirror symmetry about its armchair direction, which we represent by ${\cal M}_x$ (see Fig.~S3 of SM~\cite{Note1}). The ${\cal M}_x$ symmetry of BLG dictates that only the $\chi_{yx;xy}$ component can be finite with the $\sin2\phi$ angular dependence in $\sigma_\perp$. To explore the role of other planar Hall contributions $\propto B^2$ ($\chi_{yx;xx}$ and $\chi_{yx;yy}$), we break the ${\cal M}_x$ symmetry of BLG by applying a uniaxial strain of $1\%$ strength at an angle of $30^{\circ}$ to the zig-zag direction. The details of the strain implementation in the tight-binding model of BLG are discussed in Sec. S5 of SM~\cite{Note1}. We present the band structure of strained BLG around one of the two valleys ($K$ point) in Fig.~\ref{Fig_2}(a). In Fig.~\ref{Fig_2}(b), we show the color plot of the density of states as a function of the chemical potential ($\mu$) and inter-layer potential ($\Delta$).

For BLG, the out-of-plane position operator is given by the matrix ${\cal Z}=\frac{c}{2} \sigma_0 \otimes \tau_z$. $\sigma_0$ is the identity matrix for the sublattice space, and $\tau_z$ is the Pauli matrix in the layer space. Here, $c\approx3.35~{\rm \AA}$ is the inter-layer distance. We use this in Eqs.~\eqref{2D_BC} and \eqref{2D_OMM} to calculate the planar-BC and -OMM. We present the $x$-component of planar-BC and planar-OMM for the first conduction band in Fig.~\ref{Fig_2}(c). The $y$-components of these quantities are presented in {Sec.~S1} of SM~\cite{Note1} \footnote{We emphasize that these quantities are not ${\cal C}_{3z}$ symmetric even without any in-plane strain~\cite{Jin_prbL21, Jiang_prb20}. However, they are valley contrasting due to the global $\cal T$-symmetry of BLG.}.

We numerically calculate the 2DPHE responses of BLG, illustrating their dependence on the chemical potential $\mu$ in Fig.~\ref{Fig_2}(d). In contrast to 
$\chi_{xx;xx}$, $\chi_{xx;yy}$, and $\chi_{yx;xy}$, which remain finite even in the presence of ${\cal M}_x$ symmetry, the contributions $\chi_{xx;xy}$, $\chi_{yx;xx}$, and $\chi_{yx;yy}$ require ${\cal M}_x$ symmetry breaking to be finite. As a consequence, 
these contributions are relatively smaller in magnitude. We highlight that all the 2DPHE response tensors are pronounced in the vicinity of the band edges, where the planar band geometric quantities have a hotspot [see Fig.~\ref{Fig_2}(c)]. Interestingly, the peak in the responses near the band edge arises from the Van Hove singularity in the density of states, which is a marker of the Lifshitz transitions in BLG (see Sec. S6 of the SM~\cite{Note1} for more details).

\begin{figure}[t!]
    \centering
    \includegraphics[width = \linewidth]{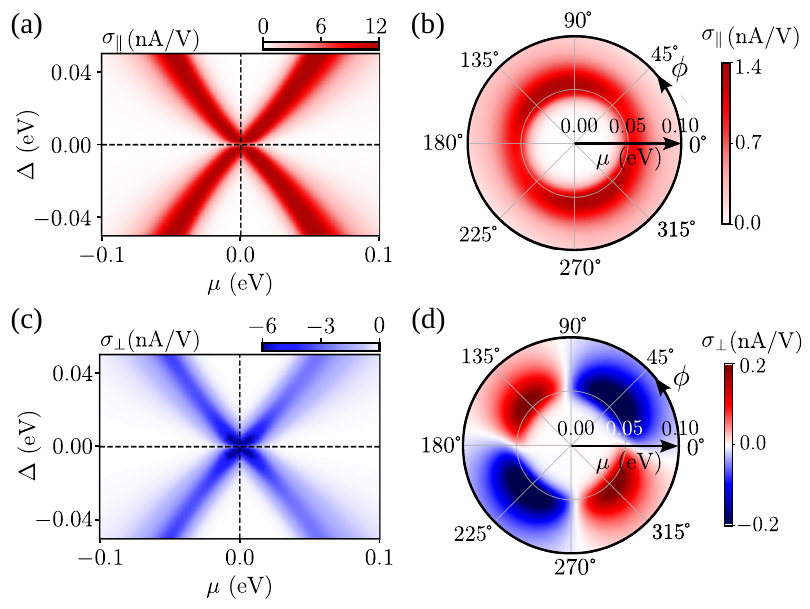}
    \caption{The color plot of the (a) longitudinal and (c) transverse 2DPHE conductivities in $\mu-\Delta$ space for strained BLG. These parameters can be experimentally tuned via the top and back gates. We have chosen the $\tau=1$ ps, $B=1$ Tesla, $\phi=60^{\circ}$ and temperature $T= 50 $K. The angular dependence of the (b) longitudinal and (d) transverse 2DPHE conductivities. The small deviation from  $\sigma_\parallel \propto \cos^2\phi$ and $\sigma_\perp \propto \sin(2\phi)$ dependence is induced by the strain-induced mirror symmetry breaking. In the angular plots, we have $\Delta = 0.05~{\rm eV}$. }
    \label{Fig_3}
\end{figure}

We present color plots of the variation of $\sigma_\parallel$ and $\sigma_\perp$ with $\mu$ and the inter-layer potential $\Delta$ in Fig.~\ref{Fig_3}(a) and ~\ref{Fig_3}(c).  
Both the conductivities have appreciable values only in the vicinity of the band edges, highlighting the band-geometric nature of 2DPHE. The peaks in both $\sigma_\perp$ and $\sigma_\parallel$ reflect the Van Hove singularity in the DOS, marked by the dark regions in Fig.~\ref{Fig_2}(b). 
We present the the angular variation of  $\sigma_\parallel$ and $\sigma_\perp$, as $\mu$ is varied, in the polar color plots in Figs.~\ref{Fig_3}(b) and ~\ref{Fig_3}(d). We highlight that the angular variation of the 2DPHE responses deviates from the conventional $\sigma_\parallel \propto \cos^2\phi$ and $\sigma_\perp \propto\sin2\phi$ dependence due to the strain-induced mirror symmetry breaking \footnote{This is a consequence of the strain-induced ${\cal M}_x$ symmetry breaking, which makes $\chi_{xx;xy}, \,\chi_{yx;xx}$ and $\chi_{yx;yy}$ components finite, modifying the angular dependence following Eqs.~\eqref{sigma_long} and \eqref{sigma_hall}. }.

{\it Experimental implications:--}
For estimating the feasibility of measuring 2DPHE responses in experiments, we consider an in-plane magnetic field of $B=1$ Tesla, $\tau = 1$ ps, and $\phi=60^{\circ}$. With these parameters, $\sigma_\perp$ ($\sigma_\parallel$) becomes $\sim 0.1 ~(4.5)~{\rm A V^{-1} m^{-1}}$ in BLG. Here, we have converted conductivities to the conventional 3D unit using the layer thickness of BLG.  
Assuming a sample size of $\sim10$ $\mu$m and a moderate electric field of $E\sim 1$ V/$\mu$m, we estimate the planar Hall voltage to be $V_\perp \sim 0.17$ $\mu$V (see Sec. S7 of SM~\cite{Note1}), which is well within experimental reach. Additionally, note that ${\bm \Omega}_{n\bm k}^{\rm planar}$ and ${\bm m}_{n\bm k}^{\rm planar}$ magnitude and hence the strength of 2DPHE responses increase with the number of layers. We show this explicitly for multi-layered graphene in Sec.~S8 of SM~\cite{Note1}. 

%

We now explore ways to distinguish 2DPHE from other possible in-plane magneto-Hall responses (see Sec. S9 of SM~\cite{Note1} for details). The corrections in the anomalous Hall velocity generated by the magnetic field-induced Berry curvature can give rise to in-plane magneto-Hall responses $\propto EB$~\cite{Zyuzin_prb20, Culcer_prl21, Carmine_prrL21, Liang_NP18, Zhou_nature22, Shengyuan22_in-plane,Wang_24layer}. The planar Hall effect can also arise from the anisotropic spin scattering mechanism~\cite{Taskin_NC17,ando_nc17, He_NP18, Wu_apl18, He_prl19, Wang_prbR20, Wang_prb21, Jinxiong_NC24}. All these in-plane magneto-Hall responses in 2D systems require strong spin-orbit coupling (SOC). Also, the responses are represented by the antisymmetric (symmetric) response tensors when they arise from the field-dependent anomalous velocity (anisotropic spin-scattering mechanism). In contrast, our proposed 2DPHE responses are symmetric and do not require SOC to be finite. Consequently, in a layered 2D system with SOC, such as transition metal dichalcogenides, the symmetrization of the total planar response will discard all antisymmetric contributions. The total symmetric planar response will have contributions from the 2DPHE and asymmetric spin scattering. However, the asymmetric spin scattering contribution is comparatively negligible in BLG due to its very small SOC strength. The asymmetric spin-scattering-induced contribution can be differentiated by its unique double-peak structure around the charge-neutrality point~\cite{ando_nc17, Wang_prbR20}. Beyond these responses, lock-in measurement techniques can easily distinguish 2DPHE from other nonlinear in-plane responses. 

{\it Conclusion:--} Our discovery of 2DPHE brings the vast class of layered 2D materials under the purview of planar Hall effect probes, which were limited to 3D materials. Additionally, 2DPHE offers a novel tool to probe the previously unexplored planar quantum-geometric properties of Bloch electrons in 2D materials. The existence of planar Berry curvature and orbital magnetic moment motivates the study of other novel phenomena, which were believed to be inaccessible in 2D materials. For instance, the planar Berry curvature can give rise to a vertical (perpendicular to the 2D plane) anomalous Hall effect in the linear and nonlinear response regimes. An interesting application of this is that vertical charge transport, with restricted out-of-plane carrier velocity, can induce an inter-layer electric polarization. This may offer a novel way to control the switching of out-of-plane electric polarization in layered ferroelectric materials~\cite{Zheng_Nat20, Wang_NatNano22, Herrero_sci21} via an in-plane electric field, 
with potential for new device applications. 2DPHE also enables the designing of highly sensitive planar Hall magnetic sensors using ultra-thin 2D  materials. Our work opens new avenues for further exploration of novel transport effects and their potential applications arising from the hidden planar band geometric quantities.  

 
\section*{Acknowledgments}
We acknowledge many fruitful discussions with Debottam Mandal (IIT Kanpur, India) and Atasi Chakraborty (Johannes Gutenberg University, Germany). KG, SD, and HV acknowledge the Ministry of Education, Government of India, for funding support through the Prime Minister's Research Fellowship program.

\bibliography{Refs}

\begin{thebibliography}{51}%
\makeatletter
\providecommand \@ifxundefined [1]{%
 \@ifx{#1\undefined}
}%
\providecommand \@ifnum [1]{%
 \ifnum #1\expandafter \@firstoftwo
 \else \expandafter \@secondoftwo
 \fi
}%
\providecommand \@ifx [1]{%
 \ifx #1\expandafter \@firstoftwo
 \else \expandafter \@secondoftwo
 \fi
}%
\providecommand \natexlab [1]{#1}%
\providecommand \enquote  [1]{``#1''}%
\providecommand \bibnamefont  [1]{#1}%
\providecommand \bibfnamefont [1]{#1}%
\providecommand \citenamefont [1]{#1}%
\providecommand \href@noop [0]{\@secondoftwo}%
\providecommand \href [0]{\begingroup \@sanitize@url \@href}%
\providecommand \@href[1]{\@@startlink{#1}\@@href}%
\providecommand \@@href[1]{\endgroup#1\@@endlink}%
\providecommand \@sanitize@url [0]{\catcode `\\12\catcode `\$12\catcode
  `\&12\catcode `\#12\catcode `\^12\catcode `\_12\catcode `\%12\relax}%
\providecommand \@@startlink[1]{}%
\providecommand \@@endlink[0]{}%
\providecommand \url  [0]{\begingroup\@sanitize@url \@url }%
\providecommand \@url [1]{\endgroup\@href {#1}{\urlprefix }}%
\providecommand \urlprefix  [0]{URL }%
\providecommand \Eprint [0]{\href }%
\providecommand \doibase [0]{https://doi.org/}%
\providecommand \selectlanguage [0]{\@gobble}%
\providecommand \bibinfo  [0]{\@secondoftwo}%
\providecommand \bibfield  [0]{\@secondoftwo}%
\providecommand \translation [1]{[#1]}%
\providecommand \BibitemOpen [0]{}%
\providecommand \bibitemStop [0]{}%
\providecommand \bibitemNoStop [0]{.\EOS\space}%
\providecommand \EOS [0]{\spacefactor3000\relax}%
\providecommand \BibitemShut  [1]{\csname bibitem#1\endcsname}%
\let\auto@bib@innerbib\@empty
\bibitem [{\citenamefont {Quynh}\ \emph {et~al.}(2019)\citenamefont {Quynh},
  \citenamefont {Hien}, \citenamefont {Binh}, \citenamefont {Dung},
  \citenamefont {Tu}, \citenamefont {Duc},\ and\ \citenamefont
  {Giang}}]{Quynh_19}%
  \BibitemOpen
  \bibfield  {author} {\bibinfo {author} {\bibfnamefont {L.~K.}\ \bibnamefont
  {Quynh}}, \bibinfo {author} {\bibfnamefont {N.~T.}\ \bibnamefont {Hien}},
  \bibinfo {author} {\bibfnamefont {N.~H.}\ \bibnamefont {Binh}}, \bibinfo
  {author} {\bibfnamefont {T.~T.}\ \bibnamefont {Dung}}, \bibinfo {author}
  {\bibfnamefont {B.~D.}\ \bibnamefont {Tu}}, \bibinfo {author} {\bibfnamefont
  {N.~H.}\ \bibnamefont {Duc}},\ and\ \bibinfo {author} {\bibfnamefont
  {D.~T.~H.}\ \bibnamefont {Giang}},\ }\bibfield  {title} {\bibinfo {title}
  {Simple planar hall effect based sensors for low-magnetic field detection},\
  }\href {https://doi.org/10.1088/2043-6254/ab1072} {\bibfield  {journal}
  {\bibinfo  {journal} {Advances in Natural Sciences: Nanoscience and
  Nanotechnology}\ }\textbf {\bibinfo {volume} {10}},\ \bibinfo {pages}
  {025002} (\bibinfo {year} {2019})}\BibitemShut {NoStop}%
\bibitem [{\citenamefont {Yau}\ and\ \citenamefont {Chang}(1971)}]{Yau_jp71}%
  \BibitemOpen
  \bibfield  {author} {\bibinfo {author} {\bibfnamefont {K.~L.}\ \bibnamefont
  {Yau}}\ and\ \bibinfo {author} {\bibfnamefont {J.~T.~H.}\ \bibnamefont
  {Chang}},\ }\bibfield  {title} {\bibinfo {title} {The planar hall effect in
  thin foils of ni-fe alloy},\ }\href
  {https://doi.org/10.1088/0305-4608/1/1/307} {\bibfield  {journal} {\bibinfo
  {journal} {Journal of Physics F: Metal Physics}\ }\textbf {\bibinfo {volume}
  {1}},\ \bibinfo {pages} {38–43} (\bibinfo {year} {1971})}\BibitemShut
  {NoStop}%
\bibitem [{\citenamefont {Li}\ \emph {et~al.}(2010)\citenamefont {Li},
  \citenamefont {Li}, \citenamefont {Wu}, \citenamefont {Li}, \citenamefont
  {Chu}, \citenamefont {Wang}, \citenamefont {Zhang}, \citenamefont {Tian},\
  and\ \citenamefont {Zheng}}]{Li_jp10}%
  \BibitemOpen
  \bibfield  {author} {\bibinfo {author} {\bibfnamefont {J.}~\bibnamefont
  {Li}}, \bibinfo {author} {\bibfnamefont {S.~L.}\ \bibnamefont {Li}}, \bibinfo
  {author} {\bibfnamefont {Z.~W.}\ \bibnamefont {Wu}}, \bibinfo {author}
  {\bibfnamefont {S.}~\bibnamefont {Li}}, \bibinfo {author} {\bibfnamefont
  {H.~F.}\ \bibnamefont {Chu}}, \bibinfo {author} {\bibfnamefont
  {J.}~\bibnamefont {Wang}}, \bibinfo {author} {\bibfnamefont {Y.}~\bibnamefont
  {Zhang}}, \bibinfo {author} {\bibfnamefont {H.~Y.}\ \bibnamefont {Tian}},\
  and\ \bibinfo {author} {\bibfnamefont {D.~N.}\ \bibnamefont {Zheng}},\
  }\bibfield  {title} {\bibinfo {title} {A phenomenological approach to the
  anisotropic magnetoresistance and planar hall effect in tetragonal
  la2/3ca1/3mno3thin films},\ }\href
  {https://doi.org/10.1088/0953-8984/22/14/146006} {\bibfield  {journal}
  {\bibinfo  {journal} {Journal of Physics: Condensed Matter}\ }\textbf
  {\bibinfo {volume} {22}},\ \bibinfo {pages} {146006} (\bibinfo {year}
  {2010})}\BibitemShut {NoStop}%
\bibitem [{\citenamefont {Tang}\ \emph {et~al.}(2003)\citenamefont {Tang},
  \citenamefont {Kawakami}, \citenamefont {Awschalom},\ and\ \citenamefont
  {Roukes}}]{Roukes_prl03}%
  \BibitemOpen
  \bibfield  {author} {\bibinfo {author} {\bibfnamefont {H.~X.}\ \bibnamefont
  {Tang}}, \bibinfo {author} {\bibfnamefont {R.~K.}\ \bibnamefont {Kawakami}},
  \bibinfo {author} {\bibfnamefont {D.~D.}\ \bibnamefont {Awschalom}},\ and\
  \bibinfo {author} {\bibfnamefont {M.~L.}\ \bibnamefont {Roukes}},\ }\bibfield
   {title} {\bibinfo {title} {Giant planar hall effect in epitaxial (ga,mn)as
  devices},\ }\href {https://doi.org/10.1103/PhysRevLett.90.107201} {\bibfield
  {journal} {\bibinfo  {journal} {Phys. Rev. Lett.}\ }\textbf {\bibinfo
  {volume} {90}},\ \bibinfo {pages} {107201} (\bibinfo {year}
  {2003})}\BibitemShut {NoStop}%
\bibitem [{\citenamefont {Bowen}\ \emph {et~al.}(2005)\citenamefont {Bowen},
  \citenamefont {Friedland}, \citenamefont {Herfort}, \citenamefont
  {Sch\"onherr},\ and\ \citenamefont {Ploog}}]{Ploong_prb05}%
  \BibitemOpen
  \bibfield  {author} {\bibinfo {author} {\bibfnamefont {M.}~\bibnamefont
  {Bowen}}, \bibinfo {author} {\bibfnamefont {K.-J.}\ \bibnamefont
  {Friedland}}, \bibinfo {author} {\bibfnamefont {J.}~\bibnamefont {Herfort}},
  \bibinfo {author} {\bibfnamefont {H.-P.}\ \bibnamefont {Sch\"onherr}},\ and\
  \bibinfo {author} {\bibfnamefont {K.~H.}\ \bibnamefont {Ploog}},\ }\bibfield
  {title} {\bibinfo {title} {Order-driven contribution to the planar hall
  effect in ${\mathrm{fe}}_{3}\mathrm{Si}$ thin films},\ }\href
  {https://doi.org/10.1103/PhysRevB.71.172401} {\bibfield  {journal} {\bibinfo
  {journal} {Phys. Rev. B}\ }\textbf {\bibinfo {volume} {71}},\ \bibinfo
  {pages} {172401} (\bibinfo {year} {2005})}\BibitemShut {NoStop}%
\bibitem [{\citenamefont {Schaefer}\ and\ \citenamefont
  {Nowack}(2021)}]{Schaefer_prb21}%
  \BibitemOpen
  \bibfield  {author} {\bibinfo {author} {\bibfnamefont {B.~T.}\ \bibnamefont
  {Schaefer}}\ and\ \bibinfo {author} {\bibfnamefont {K.~C.}\ \bibnamefont
  {Nowack}},\ }\bibfield  {title} {\bibinfo {title} {Electrically tunable and
  reversible magnetoelectric coupling in strained bilayer graphene},\ }\href
  {https://doi.org/10.1103/PhysRevB.103.224426} {\bibfield  {journal} {\bibinfo
   {journal} {Phys. Rev. B}\ }\textbf {\bibinfo {volume} {103}},\ \bibinfo
  {pages} {224426} (\bibinfo {year} {2021})}\BibitemShut {NoStop}%
\bibitem [{\citenamefont {Nandy}\ \emph {et~al.}(2017)\citenamefont {Nandy},
  \citenamefont {Sharma}, \citenamefont {Taraphder},\ and\ \citenamefont
  {Tewari}}]{Nandy_prl17}%
  \BibitemOpen
  \bibfield  {author} {\bibinfo {author} {\bibfnamefont {S.}~\bibnamefont
  {Nandy}}, \bibinfo {author} {\bibfnamefont {G.}~\bibnamefont {Sharma}},
  \bibinfo {author} {\bibfnamefont {A.}~\bibnamefont {Taraphder}},\ and\
  \bibinfo {author} {\bibfnamefont {S.}~\bibnamefont {Tewari}},\ }\bibfield
  {title} {\bibinfo {title} {Chiral anomaly as the origin of the planar hall
  effect in weyl semimetals},\ }\href
  {https://doi.org/10.1103/PhysRevLett.119.176804} {\bibfield  {journal}
  {\bibinfo  {journal} {Phys. Rev. Lett.}\ }\textbf {\bibinfo {volume} {119}},\
  \bibinfo {pages} {176804} (\bibinfo {year} {2017})}\BibitemShut {NoStop}%
\bibitem [{\citenamefont {Kumar}\ \emph {et~al.}(2018)\citenamefont {Kumar},
  \citenamefont {Guin}, \citenamefont {Felser},\ and\ \citenamefont
  {Shekhar}}]{shekhar_prb18}%
  \BibitemOpen
  \bibfield  {author} {\bibinfo {author} {\bibfnamefont {N.}~\bibnamefont
  {Kumar}}, \bibinfo {author} {\bibfnamefont {S.~N.}\ \bibnamefont {Guin}},
  \bibinfo {author} {\bibfnamefont {C.}~\bibnamefont {Felser}},\ and\ \bibinfo
  {author} {\bibfnamefont {C.}~\bibnamefont {Shekhar}},\ }\bibfield  {title}
  {\bibinfo {title} {Planar hall effect in the weyl semimetal gdptbi},\ }\href
  {https://doi.org/10.1103/PhysRevB.98.041103} {\bibfield  {journal} {\bibinfo
  {journal} {Phys. Rev. B}\ }\textbf {\bibinfo {volume} {98}},\ \bibinfo
  {pages} {041103} (\bibinfo {year} {2018})}\BibitemShut {NoStop}%
\bibitem [{\citenamefont {Deng}\ \emph {et~al.}(2019)\citenamefont {Deng},
  \citenamefont {Duan}, \citenamefont {Luo}, \citenamefont {Deng},
  \citenamefont {Wang},\ and\ \citenamefont {Sheng}}]{deng_prb19}%
  \BibitemOpen
  \bibfield  {author} {\bibinfo {author} {\bibfnamefont {M.-X.}\ \bibnamefont
  {Deng}}, \bibinfo {author} {\bibfnamefont {H.-J.}\ \bibnamefont {Duan}},
  \bibinfo {author} {\bibfnamefont {W.}~\bibnamefont {Luo}}, \bibinfo {author}
  {\bibfnamefont {W.~Y.}\ \bibnamefont {Deng}}, \bibinfo {author}
  {\bibfnamefont {R.-Q.}\ \bibnamefont {Wang}},\ and\ \bibinfo {author}
  {\bibfnamefont {L.}~\bibnamefont {Sheng}},\ }\bibfield  {title} {\bibinfo
  {title} {Quantum oscillation modulated angular dependence of the positive
  longitudinal magnetoconductivity and planar hall effect in weyl semimetals},\
  }\href {https://doi.org/10.1103/PhysRevB.99.165146} {\bibfield  {journal}
  {\bibinfo  {journal} {Phys. Rev. B}\ }\textbf {\bibinfo {volume} {99}},\
  \bibinfo {pages} {165146} (\bibinfo {year} {2019})}\BibitemShut {NoStop}%
\bibitem [{\citenamefont {Li}\ \emph {et~al.}(2023)\citenamefont {Li},
  \citenamefont {Cao}, \citenamefont {Cui}, \citenamefont {Yu},\ and\
  \citenamefont {Yao}}]{Yao_prb23}%
  \BibitemOpen
  \bibfield  {author} {\bibinfo {author} {\bibfnamefont {L.}~\bibnamefont
  {Li}}, \bibinfo {author} {\bibfnamefont {J.}~\bibnamefont {Cao}}, \bibinfo
  {author} {\bibfnamefont {C.}~\bibnamefont {Cui}}, \bibinfo {author}
  {\bibfnamefont {Z.-M.}\ \bibnamefont {Yu}},\ and\ \bibinfo {author}
  {\bibfnamefont {Y.}~\bibnamefont {Yao}},\ }\bibfield  {title} {\bibinfo
  {title} {Planar hall effect in topological weyl and nodal-line semimetals},\
  }\href {https://doi.org/10.1103/PhysRevB.108.085120} {\bibfield  {journal}
  {\bibinfo  {journal} {Phys. Rev. B}\ }\textbf {\bibinfo {volume} {108}},\
  \bibinfo {pages} {085120} (\bibinfo {year} {2023})}\BibitemShut {NoStop}%
\bibitem [{\citenamefont {Ma}\ \emph {et~al.}(2019)\citenamefont {Ma},
  \citenamefont {Jiang}, \citenamefont {Liu},\ and\ \citenamefont
  {Xie}}]{Xie_prb19}%
  \BibitemOpen
  \bibfield  {author} {\bibinfo {author} {\bibfnamefont {D.}~\bibnamefont
  {Ma}}, \bibinfo {author} {\bibfnamefont {H.}~\bibnamefont {Jiang}}, \bibinfo
  {author} {\bibfnamefont {H.}~\bibnamefont {Liu}},\ and\ \bibinfo {author}
  {\bibfnamefont {X.~C.}\ \bibnamefont {Xie}},\ }\bibfield  {title} {\bibinfo
  {title} {Planar hall effect in tilted weyl semimetals},\ }\href
  {https://doi.org/10.1103/PhysRevB.99.115121} {\bibfield  {journal} {\bibinfo
  {journal} {Phys. Rev. B}\ }\textbf {\bibinfo {volume} {99}},\ \bibinfo
  {pages} {115121} (\bibinfo {year} {2019})}\BibitemShut {NoStop}%
\bibitem [{\citenamefont {Das}\ \emph {et~al.}(2020)\citenamefont {Das},
  \citenamefont {Singh},\ and\ \citenamefont {Agarwal}}]{Kamal_prr20_chiral}%
  \BibitemOpen
  \bibfield  {author} {\bibinfo {author} {\bibfnamefont {K.}~\bibnamefont
  {Das}}, \bibinfo {author} {\bibfnamefont {S.~K.}\ \bibnamefont {Singh}},\
  and\ \bibinfo {author} {\bibfnamefont {A.}~\bibnamefont {Agarwal}},\
  }\bibfield  {title} {\bibinfo {title} {Chiral anomalies induced transport in
  weyl metals in quantizing magnetic field},\ }\href
  {https://doi.org/10.1103/PhysRevResearch.2.033511} {\bibfield  {journal}
  {\bibinfo  {journal} {Phys. Rev. Res.}\ }\textbf {\bibinfo {volume} {2}},\
  \bibinfo {pages} {033511} (\bibinfo {year} {2020})}\BibitemShut {NoStop}%
\bibitem [{\citenamefont {Wei}\ \emph {et~al.}(2023)\citenamefont {Wei},
  \citenamefont {Feng},\ and\ \citenamefont {Weng}}]{Weng_prb23}%
  \BibitemOpen
  \bibfield  {author} {\bibinfo {author} {\bibfnamefont {Y.-W.}\ \bibnamefont
  {Wei}}, \bibinfo {author} {\bibfnamefont {J.}~\bibnamefont {Feng}},\ and\
  \bibinfo {author} {\bibfnamefont {H.}~\bibnamefont {Weng}},\ }\bibfield
  {title} {\bibinfo {title} {Spatial symmetry modulation of planar hall effect
  in weyl semimetals},\ }\href {https://doi.org/10.1103/PhysRevB.107.075131}
  {\bibfield  {journal} {\bibinfo  {journal} {Phys. Rev. B}\ }\textbf {\bibinfo
  {volume} {107}},\ \bibinfo {pages} {075131} (\bibinfo {year}
  {2023})}\BibitemShut {NoStop}%
\bibitem [{\citenamefont {Das}\ and\ \citenamefont
  {Agarwal}(2019)}]{Kamal_prb19}%
  \BibitemOpen
  \bibfield  {author} {\bibinfo {author} {\bibfnamefont {K.}~\bibnamefont
  {Das}}\ and\ \bibinfo {author} {\bibfnamefont {A.}~\bibnamefont {Agarwal}},\
  }\bibfield  {title} {\bibinfo {title} {Linear magnetochiral transport in
  tilted type-i and type-ii weyl semimetals},\ }\href
  {https://doi.org/10.1103/PhysRevB.99.085405} {\bibfield  {journal} {\bibinfo
  {journal} {Phys. Rev. B}\ }\textbf {\bibinfo {volume} {99}},\ \bibinfo
  {pages} {085405} (\bibinfo {year} {2019})}\BibitemShut {NoStop}%
\bibitem [{\citenamefont {Das}\ \emph {et~al.}(2023)\citenamefont {Das},
  \citenamefont {Das},\ and\ \citenamefont {Agarwal}}]{Sunit_prb23}%
  \BibitemOpen
  \bibfield  {author} {\bibinfo {author} {\bibfnamefont {S.}~\bibnamefont
  {Das}}, \bibinfo {author} {\bibfnamefont {K.}~\bibnamefont {Das}},\ and\
  \bibinfo {author} {\bibfnamefont {A.}~\bibnamefont {Agarwal}},\ }\bibfield
  {title} {\bibinfo {title} {Chiral anomalies in three-dimensional spin-orbit
  coupled metals: Electrical, thermal, and gravitational anomalies},\ }\href
  {https://doi.org/10.1103/PhysRevB.108.045405} {\bibfield  {journal} {\bibinfo
   {journal} {Phys. Rev. B}\ }\textbf {\bibinfo {volume} {108}},\ \bibinfo
  {pages} {045405} (\bibinfo {year} {2023})}\BibitemShut {NoStop}%
\bibitem [{\citenamefont {Ghosh}\ and\ \citenamefont
  {Mandal}(2024{\natexlab{a}})}]{Ipsita_24}%
  \BibitemOpen
  \bibfield  {author} {\bibinfo {author} {\bibfnamefont {R.}~\bibnamefont
  {Ghosh}}\ and\ \bibinfo {author} {\bibfnamefont {I.}~\bibnamefont {Mandal}},\
  }\bibfield  {title} {\bibinfo {title} {Direction-dependent conductivity in
  planar hall set-ups with tilted weyl/multi-weyl semimetals},\ }\href
  {https://doi.org/10.1088/1361-648x/ad38fa} {\bibfield  {journal} {\bibinfo
  {journal} {Journal of Physics: Condensed Matter}\ }\textbf {\bibinfo {volume}
  {36}},\ \bibinfo {pages} {275501} (\bibinfo {year}
  {2024}{\natexlab{a}})}\BibitemShut {NoStop}%
\bibitem [{\citenamefont {Ghosh}\ and\ \citenamefont
  {Mandal}(2024{\natexlab{b}})}]{Ipsita_physica24}%
  \BibitemOpen
  \bibfield  {author} {\bibinfo {author} {\bibfnamefont {R.}~\bibnamefont
  {Ghosh}}\ and\ \bibinfo {author} {\bibfnamefont {I.}~\bibnamefont {Mandal}},\
  }\bibfield  {title} {\bibinfo {title} {Electric and thermoelectric response
  for weyl and multi-weyl semimetals in planar hall configurations including
  the effects of strain},\ }\href {https://doi.org/10.1016/j.physe.2024.115914}
  {\bibfield  {journal} {\bibinfo  {journal} {Physica E: Low-dimensional
  Systems and Nanostructures}\ }\textbf {\bibinfo {volume} {159}},\ \bibinfo
  {pages} {115914} (\bibinfo {year} {2024}{\natexlab{b}})}\BibitemShut
  {NoStop}%
\bibitem [{\citenamefont {Rakhmilevich}\ \emph {et~al.}(2018)\citenamefont
  {Rakhmilevich}, \citenamefont {Wang}, \citenamefont {Zhao}, \citenamefont
  {Chan}, \citenamefont {Moodera}, \citenamefont {Liu},\ and\ \citenamefont
  {Chang}}]{Chang_prb18}%
  \BibitemOpen
  \bibfield  {author} {\bibinfo {author} {\bibfnamefont {D.}~\bibnamefont
  {Rakhmilevich}}, \bibinfo {author} {\bibfnamefont {F.}~\bibnamefont {Wang}},
  \bibinfo {author} {\bibfnamefont {W.}~\bibnamefont {Zhao}}, \bibinfo {author}
  {\bibfnamefont {M.~H.~W.}\ \bibnamefont {Chan}}, \bibinfo {author}
  {\bibfnamefont {J.~S.}\ \bibnamefont {Moodera}}, \bibinfo {author}
  {\bibfnamefont {C.}~\bibnamefont {Liu}},\ and\ \bibinfo {author}
  {\bibfnamefont {C.-Z.}\ \bibnamefont {Chang}},\ }\bibfield  {title} {\bibinfo
  {title} {Unconventional planar hall effect in exchange-coupled topological
  insulator--ferromagnetic insulator heterostructures},\ }\href
  {https://doi.org/10.1103/PhysRevB.98.094404} {\bibfield  {journal} {\bibinfo
  {journal} {Phys. Rev. B}\ }\textbf {\bibinfo {volume} {98}},\ \bibinfo
  {pages} {094404} (\bibinfo {year} {2018})}\BibitemShut {NoStop}%
\bibitem [{\citenamefont {Nandy}\ \emph {et~al.}(2018)\citenamefont {Nandy},
  \citenamefont {Taraphder},\ and\ \citenamefont {Tewari}}]{Nandy_sr18}%
  \BibitemOpen
  \bibfield  {author} {\bibinfo {author} {\bibfnamefont {S.}~\bibnamefont
  {Nandy}}, \bibinfo {author} {\bibfnamefont {A.}~\bibnamefont {Taraphder}},\
  and\ \bibinfo {author} {\bibfnamefont {S.}~\bibnamefont {Tewari}},\
  }\bibfield  {title} {\bibinfo {title} {Berry phase theory of planar hall
  effect in topological insulators},\ }\href
  {https://doi.org/10.1038/s41598-018-33258-5} {\bibfield  {journal} {\bibinfo
  {journal} {Scientific Reports}\ }\textbf {\bibinfo {volume} {8}},\ \bibinfo
  {pages} {14983} (\bibinfo {year} {2018})}\BibitemShut {NoStop}%
\bibitem [{\citenamefont {Xiao}\ \emph {et~al.}(2010)\citenamefont {Xiao},
  \citenamefont {Chang},\ and\ \citenamefont {Niu}}]{Xiao_rmp10}%
  \BibitemOpen
  \bibfield  {author} {\bibinfo {author} {\bibfnamefont {D.}~\bibnamefont
  {Xiao}}, \bibinfo {author} {\bibfnamefont {M.-C.}\ \bibnamefont {Chang}},\
  and\ \bibinfo {author} {\bibfnamefont {Q.}~\bibnamefont {Niu}},\ }\bibfield
  {title} {\bibinfo {title} {Berry phase effects on electronic properties},\
  }\href {https://link.aps.org/doi/10.1103/RevModPhys.82.1959} {\bibfield
  {journal} {\bibinfo  {journal} {Rev. Mod. Phys.}\ }\textbf {\bibinfo {volume}
  {82}},\ \bibinfo {pages} {1959} (\bibinfo {year} {2010})}\BibitemShut
  {NoStop}%
\bibitem [{\citenamefont {Cayssol}\ and\ \citenamefont
  {Fuchs}(2021)}]{Cayssol_JPM21}%
  \BibitemOpen
  \bibfield  {author} {\bibinfo {author} {\bibfnamefont {J.}~\bibnamefont
  {Cayssol}}\ and\ \bibinfo {author} {\bibfnamefont {J.~N.}\ \bibnamefont
  {Fuchs}},\ }\bibfield  {title} {\bibinfo {title} {Topological and geometrical
  aspects of band theory},\ }\href {https://doi.org/10.1088/2515-7639/abf0b5}
  {\bibfield  {journal} {\bibinfo  {journal} {Journal of Physics: Materials}\
  }\textbf {\bibinfo {volume} {4}},\ \bibinfo {pages} {034007} (\bibinfo {year}
  {2021})}\BibitemShut {NoStop}%
\bibitem [{\citenamefont {Zyuzin}(2020)}]{Zyuzin_prb20}%
  \BibitemOpen
  \bibfield  {author} {\bibinfo {author} {\bibfnamefont {V.~A.}\ \bibnamefont
  {Zyuzin}},\ }\bibfield  {title} {\bibinfo {title} {In-plane hall effect in
  two-dimensional helical electron systems},\ }\href
  {https://doi.org/10.1103/PhysRevB.102.241105} {\bibfield  {journal} {\bibinfo
   {journal} {Phys. Rev. B}\ }\textbf {\bibinfo {volume} {102}},\ \bibinfo
  {pages} {241105} (\bibinfo {year} {2020})}\BibitemShut {NoStop}%
\bibitem [{\citenamefont {Cullen}\ \emph {et~al.}(2021)\citenamefont {Cullen},
  \citenamefont {Bhalla}, \citenamefont {Marcellina}, \citenamefont
  {Hamilton},\ and\ \citenamefont {Culcer}}]{Culcer_prl21}%
  \BibitemOpen
  \bibfield  {author} {\bibinfo {author} {\bibfnamefont {J.~H.}\ \bibnamefont
  {Cullen}}, \bibinfo {author} {\bibfnamefont {P.}~\bibnamefont {Bhalla}},
  \bibinfo {author} {\bibfnamefont {E.}~\bibnamefont {Marcellina}}, \bibinfo
  {author} {\bibfnamefont {A.~R.}\ \bibnamefont {Hamilton}},\ and\ \bibinfo
  {author} {\bibfnamefont {D.}~\bibnamefont {Culcer}},\ }\bibfield  {title}
  {\bibinfo {title} {Generating a topological anomalous hall effect in a
  nonmagnetic conductor: An in-plane magnetic field as a direct probe of the
  berry curvature},\ }\href {https://doi.org/10.1103/PhysRevLett.126.256601}
  {\bibfield  {journal} {\bibinfo  {journal} {Phys. Rev. Lett.}\ }\textbf
  {\bibinfo {volume} {126}},\ \bibinfo {pages} {256601} (\bibinfo {year}
  {2021})}\BibitemShut {NoStop}%
\bibitem [{\citenamefont {Battilomo}\ \emph {et~al.}(2021)\citenamefont
  {Battilomo}, \citenamefont {Scopigno},\ and\ \citenamefont
  {Ortix}}]{Carmine_prrL21}%
  \BibitemOpen
  \bibfield  {author} {\bibinfo {author} {\bibfnamefont {R.}~\bibnamefont
  {Battilomo}}, \bibinfo {author} {\bibfnamefont {N.}~\bibnamefont
  {Scopigno}},\ and\ \bibinfo {author} {\bibfnamefont {C.}~\bibnamefont
  {Ortix}},\ }\bibfield  {title} {\bibinfo {title} {Anomalous planar hall
  effect in two-dimensional trigonal crystals},\ }\href
  {https://doi.org/10.1103/PhysRevResearch.3.L012006} {\bibfield  {journal}
  {\bibinfo  {journal} {Phys. Rev. Res.}\ }\textbf {\bibinfo {volume} {3}},\
  \bibinfo {pages} {L012006} (\bibinfo {year} {2021})}\BibitemShut {NoStop}%
\bibitem [{\citenamefont {Liang}\ \emph {et~al.}(2018)\citenamefont {Liang},
  \citenamefont {Lin}, \citenamefont {Gibson}, \citenamefont {Kushwaha},
  \citenamefont {Liu}, \citenamefont {Wang}, \citenamefont {Xiong},
  \citenamefont {Sobota}, \citenamefont {Hashimoto}, \citenamefont {Kirchmann},
  \citenamefont {Shen}, \citenamefont {Cava},\ and\ \citenamefont
  {Ong}}]{Liang_NP18}%
  \BibitemOpen
  \bibfield  {author} {\bibinfo {author} {\bibfnamefont {T.}~\bibnamefont
  {Liang}}, \bibinfo {author} {\bibfnamefont {J.}~\bibnamefont {Lin}}, \bibinfo
  {author} {\bibfnamefont {Q.}~\bibnamefont {Gibson}}, \bibinfo {author}
  {\bibfnamefont {S.}~\bibnamefont {Kushwaha}}, \bibinfo {author}
  {\bibfnamefont {M.}~\bibnamefont {Liu}}, \bibinfo {author} {\bibfnamefont
  {W.}~\bibnamefont {Wang}}, \bibinfo {author} {\bibfnamefont {H.}~\bibnamefont
  {Xiong}}, \bibinfo {author} {\bibfnamefont {J.~A.}\ \bibnamefont {Sobota}},
  \bibinfo {author} {\bibfnamefont {M.}~\bibnamefont {Hashimoto}}, \bibinfo
  {author} {\bibfnamefont {P.~S.}\ \bibnamefont {Kirchmann}}, \bibinfo {author}
  {\bibfnamefont {Z.-X.}\ \bibnamefont {Shen}}, \bibinfo {author}
  {\bibfnamefont {R.~J.}\ \bibnamefont {Cava}},\ and\ \bibinfo {author}
  {\bibfnamefont {N.~P.}\ \bibnamefont {Ong}},\ }\bibfield  {title} {\bibinfo
  {title} {Anomalous hall effect in {ZrTe}5},\ }\href
  {https://doi.org/10.1038/s41567-018-0078-z} {\bibfield  {journal} {\bibinfo
  {journal} {Nature Physics}\ }\textbf {\bibinfo {volume} {14}},\ \bibinfo
  {pages} {451} (\bibinfo {year} {2018})}\BibitemShut {NoStop}%
\bibitem [{\citenamefont {Zhou}\ \emph {et~al.}(2022)\citenamefont {Zhou},
  \citenamefont {Zhang}, \citenamefont {Lin}, \citenamefont {Cao},
  \citenamefont {Zhou}, \citenamefont {Jiang}, \citenamefont {Du},
  \citenamefont {Tang}, \citenamefont {Shi}, \citenamefont {Jiang},
  \citenamefont {Cao}, \citenamefont {Lin}, \citenamefont {Fu}, \citenamefont
  {Zhu}, \citenamefont {Guo}, \citenamefont {Huang}, \citenamefont {Yao},
  \citenamefont {Parkin}, \citenamefont {Zhou}, \citenamefont {Gao},
  \citenamefont {Wang}, \citenamefont {Hou}, \citenamefont {Yao}, \citenamefont
  {Suenaga}, \citenamefont {Wu},\ and\ \citenamefont {Liu}}]{Zhou_nature22}%
  \BibitemOpen
  \bibfield  {author} {\bibinfo {author} {\bibfnamefont {J.}~\bibnamefont
  {Zhou}}, \bibinfo {author} {\bibfnamefont {W.}~\bibnamefont {Zhang}},
  \bibinfo {author} {\bibfnamefont {Y.-C.}\ \bibnamefont {Lin}}, \bibinfo
  {author} {\bibfnamefont {J.}~\bibnamefont {Cao}}, \bibinfo {author}
  {\bibfnamefont {Y.}~\bibnamefont {Zhou}}, \bibinfo {author} {\bibfnamefont
  {W.}~\bibnamefont {Jiang}}, \bibinfo {author} {\bibfnamefont
  {H.}~\bibnamefont {Du}}, \bibinfo {author} {\bibfnamefont {B.}~\bibnamefont
  {Tang}}, \bibinfo {author} {\bibfnamefont {J.}~\bibnamefont {Shi}}, \bibinfo
  {author} {\bibfnamefont {B.}~\bibnamefont {Jiang}}, \bibinfo {author}
  {\bibfnamefont {X.}~\bibnamefont {Cao}}, \bibinfo {author} {\bibfnamefont
  {B.}~\bibnamefont {Lin}}, \bibinfo {author} {\bibfnamefont {Q.}~\bibnamefont
  {Fu}}, \bibinfo {author} {\bibfnamefont {C.}~\bibnamefont {Zhu}}, \bibinfo
  {author} {\bibfnamefont {W.}~\bibnamefont {Guo}}, \bibinfo {author}
  {\bibfnamefont {Y.}~\bibnamefont {Huang}}, \bibinfo {author} {\bibfnamefont
  {Y.}~\bibnamefont {Yao}}, \bibinfo {author} {\bibfnamefont {S.~S.~P.}\
  \bibnamefont {Parkin}}, \bibinfo {author} {\bibfnamefont {J.}~\bibnamefont
  {Zhou}}, \bibinfo {author} {\bibfnamefont {Y.}~\bibnamefont {Gao}}, \bibinfo
  {author} {\bibfnamefont {Y.}~\bibnamefont {Wang}}, \bibinfo {author}
  {\bibfnamefont {Y.}~\bibnamefont {Hou}}, \bibinfo {author} {\bibfnamefont
  {Y.}~\bibnamefont {Yao}}, \bibinfo {author} {\bibfnamefont {K.}~\bibnamefont
  {Suenaga}}, \bibinfo {author} {\bibfnamefont {X.}~\bibnamefont {Wu}},\ and\
  \bibinfo {author} {\bibfnamefont {Z.}~\bibnamefont {Liu}},\ }\bibfield
  {title} {\bibinfo {title} {Heterodimensional superlattice with
  in-plane~anomalous hall effect},\ }\href
  {https://doi.org/10.1038/s41586-022-05031-2} {\bibfield  {journal} {\bibinfo
  {journal} {Nature}\ }\textbf {\bibinfo {volume} {609}},\ \bibinfo {pages}
  {46} (\bibinfo {year} {2022})}\BibitemShut {NoStop}%
\bibitem [{\citenamefont {Sun}\ \emph {et~al.}(2022)\citenamefont {Sun},
  \citenamefont {Weng},\ and\ \citenamefont {Dai}}]{Dai_prbL22}%
  \BibitemOpen
  \bibfield  {author} {\bibinfo {author} {\bibfnamefont {S.}~\bibnamefont
  {Sun}}, \bibinfo {author} {\bibfnamefont {H.}~\bibnamefont {Weng}},\ and\
  \bibinfo {author} {\bibfnamefont {X.}~\bibnamefont {Dai}},\ }\bibfield
  {title} {\bibinfo {title} {Possible quantization and half-quantization in the
  anomalous hall effect caused by in-plane magnetic field},\ }\href
  {https://doi.org/10.1103/PhysRevB.106.L241105} {\bibfield  {journal}
  {\bibinfo  {journal} {Phys. Rev. B}\ }\textbf {\bibinfo {volume} {106}},\
  \bibinfo {pages} {L241105} (\bibinfo {year} {2022})}\BibitemShut {NoStop}%
\bibitem [{\citenamefont {Wang}\ \emph {et~al.}(2024)\citenamefont {Wang},
  \citenamefont {Huang}, \citenamefont {Liu}, \citenamefont {Feng},
  \citenamefont {Zhu}, \citenamefont {Wu}, \citenamefont {Xiao},\ and\
  \citenamefont {Yang}}]{Shengyuan22_in-plane}%
  \BibitemOpen
  \bibfield  {author} {\bibinfo {author} {\bibfnamefont {H.}~\bibnamefont
  {Wang}}, \bibinfo {author} {\bibfnamefont {Y.-X.}\ \bibnamefont {Huang}},
  \bibinfo {author} {\bibfnamefont {H.}~\bibnamefont {Liu}}, \bibinfo {author}
  {\bibfnamefont {X.}~\bibnamefont {Feng}}, \bibinfo {author} {\bibfnamefont
  {J.}~\bibnamefont {Zhu}}, \bibinfo {author} {\bibfnamefont {W.}~\bibnamefont
  {Wu}}, \bibinfo {author} {\bibfnamefont {C.}~\bibnamefont {Xiao}},\ and\
  \bibinfo {author} {\bibfnamefont {S.~A.}\ \bibnamefont {Yang}},\ }\bibfield
  {title} {\bibinfo {title} {Orbital origin of the intrinsic planar hall
  effect},\ }\href {https://doi.org/10.1103/PhysRevLett.132.056301} {\bibfield
  {journal} {\bibinfo  {journal} {Phys. Rev. Lett.}\ }\textbf {\bibinfo
  {volume} {132}},\ \bibinfo {pages} {056301} (\bibinfo {year}
  {2024})}\BibitemShut {NoStop}%
\bibitem [{\citenamefont {Drigo}\ and\ \citenamefont
  {Resta}(2020)}]{Resta_prb20}%
  \BibitemOpen
  \bibfield  {author} {\bibinfo {author} {\bibfnamefont {E.}~\bibnamefont
  {Drigo}}\ and\ \bibinfo {author} {\bibfnamefont {R.}~\bibnamefont {Resta}},\
  }\bibfield  {title} {\bibinfo {title} {Chern number and orbital magnetization
  in ribbons, polymers, and single-layer materials},\ }\href
  {https://doi.org/10.1103/PhysRevB.101.165120} {\bibfield  {journal} {\bibinfo
   {journal} {Phys. Rev. B}\ }\textbf {\bibinfo {volume} {101}},\ \bibinfo
  {pages} {165120} (\bibinfo {year} {2020})}\BibitemShut {NoStop}%
\bibitem [{\citenamefont {Hara}\ \emph {et~al.}(2020)\citenamefont {Hara},
  \citenamefont {Bahramy},\ and\ \citenamefont {Murakami}}]{murakami_prb20}%
  \BibitemOpen
  \bibfield  {author} {\bibinfo {author} {\bibfnamefont {D.}~\bibnamefont
  {Hara}}, \bibinfo {author} {\bibfnamefont {M.~S.}\ \bibnamefont {Bahramy}},\
  and\ \bibinfo {author} {\bibfnamefont {S.}~\bibnamefont {Murakami}},\
  }\bibfield  {title} {\bibinfo {title} {Current-induced orbital magnetization
  in systems without inversion symmetry},\ }\href
  {https://doi.org/10.1103/PhysRevB.102.184404} {\bibfield  {journal} {\bibinfo
   {journal} {Phys. Rev. B}\ }\textbf {\bibinfo {volume} {102}},\ \bibinfo
  {pages} {184404} (\bibinfo {year} {2020})}\BibitemShut {NoStop}%
\bibitem [{\citenamefont {Kim}\ \emph {et~al.}(2021)\citenamefont {Kim},
  \citenamefont {Jeong}, \citenamefont {Kim},\ and\ \citenamefont
  {Jin}}]{Jin_prbL21}%
  \BibitemOpen
  \bibfield  {author} {\bibinfo {author} {\bibfnamefont {K.-W.}\ \bibnamefont
  {Kim}}, \bibinfo {author} {\bibfnamefont {H.}~\bibnamefont {Jeong}}, \bibinfo
  {author} {\bibfnamefont {J.}~\bibnamefont {Kim}},\ and\ \bibinfo {author}
  {\bibfnamefont {H.}~\bibnamefont {Jin}},\ }\bibfield  {title} {\bibinfo
  {title} {Vertical transverse transport induced by hidden in-plane berry
  curvature in two dimensions},\ }\href
  {https://doi.org/10.1103/PhysRevB.104.L081114} {\bibfield  {journal}
  {\bibinfo  {journal} {Phys. Rev. B}\ }\textbf {\bibinfo {volume} {104}},\
  \bibinfo {pages} {L081114} (\bibinfo {year} {2021})}\BibitemShut {NoStop}%
\bibitem [{\citenamefont {Duong}\ \emph {et~al.}(2017)\citenamefont {Duong},
  \citenamefont {Yun},\ and\ \citenamefont {Lee}}]{Duong_ACS17_vdW}%
  \BibitemOpen
  \bibfield  {author} {\bibinfo {author} {\bibfnamefont {D.~L.}\ \bibnamefont
  {Duong}}, \bibinfo {author} {\bibfnamefont {S.~J.}\ \bibnamefont {Yun}},\
  and\ \bibinfo {author} {\bibfnamefont {Y.~H.}\ \bibnamefont {Lee}},\
  }\bibfield  {title} {\bibinfo {title} {van der waals layered materials:
  Opportunities and challenges},\ }\href
  {https://doi.org/10.1021/acsnano.7b07436} {\bibfield  {journal} {\bibinfo
  {journal} {ACS Nano}\ }\textbf {\bibinfo {volume} {11}},\ \bibinfo {pages}
  {11803–11830} (\bibinfo {year} {2017})}\BibitemShut {NoStop}%
\bibitem [{Note1()}]{Note1}%
  \BibitemOpen
  \bibinfo {note} {The \protect \href
  {https://www.dropbox.com/scl/fi/fxb532ylx6dk5xicrcgea/SM_2DPHE.pdf?rlkey=r2ajvbgg2j6t5y3z4xbwihw3u&st=9av0zfu2&dl=0}{Supplementary
  Material} discusses: i) the derivation of planar-BC and planar-OMM
  expressions, ii) general expression for planar-BC and planar-OMM and
  analytical calculation of them for $2\times 2$ low-energy bilayer graphene
  model. iii) the detailed derivation of longitudinal and planar Hall response
  tensors, iv) the details of symmetry analysis, v) the strain implementation
  in the tight-binding model for bilayer graphene and $y$ components of planar
  Berry curvature and OMM, vi) the Van Hove singularity and Lifshitz transition
  of Fermi surface, vii) estimation of planar Hall voltage, viii) comparison of
  planar Berry curvature and OMM for bilayer and trilayer graphene, and ix)
  other in-plane magneto-Hall responses in two-dimensional
  systems.}\BibitemShut {Stop}%
\bibitem [{\citenamefont {Ho}\ \emph {et~al.}(2021)\citenamefont {Ho},
  \citenamefont {Chang}, \citenamefont {Hsieh}, \citenamefont {Lo},
  \citenamefont {Huang}, \citenamefont {Vu}, \citenamefont {Ortix},\ and\
  \citenamefont {Chen}}]{Chen_NE21}%
  \BibitemOpen
  \bibfield  {author} {\bibinfo {author} {\bibfnamefont {S.-C.}\ \bibnamefont
  {Ho}}, \bibinfo {author} {\bibfnamefont {C.-H.}\ \bibnamefont {Chang}},
  \bibinfo {author} {\bibfnamefont {Y.-C.}\ \bibnamefont {Hsieh}}, \bibinfo
  {author} {\bibfnamefont {S.-T.}\ \bibnamefont {Lo}}, \bibinfo {author}
  {\bibfnamefont {B.}~\bibnamefont {Huang}}, \bibinfo {author} {\bibfnamefont
  {T.-H.-Y.}\ \bibnamefont {Vu}}, \bibinfo {author} {\bibfnamefont
  {C.}~\bibnamefont {Ortix}},\ and\ \bibinfo {author} {\bibfnamefont {T.-M.}\
  \bibnamefont {Chen}},\ }\bibfield  {title} {\bibinfo {title} {Hall effects in
  artificially corrugated bilayer graphene without breaking time-reversal
  symmetry},\ }\href {https://doi.org/10.1038/s41928-021-00537-5} {\bibfield
  {journal} {\bibinfo  {journal} {Nature Electronics}\ }\textbf {\bibinfo
  {volume} {4}},\ \bibinfo {pages} {116} (\bibinfo {year} {2021})}\BibitemShut
  {NoStop}%
\bibitem [{\citenamefont {Das}\ and\ \citenamefont
  {Agarwal}(2020)}]{Das_PRR_2020a}%
  \BibitemOpen
  \bibfield  {author} {\bibinfo {author} {\bibfnamefont {K.}~\bibnamefont
  {Das}}\ and\ \bibinfo {author} {\bibfnamefont {A.}~\bibnamefont {Agarwal}},\
  }\bibfield  {title} {\bibinfo {title} {Thermal and gravitational chiral
  anomaly induced magneto-transport in weyl semimetals},\ }\href
  {https://doi.org/10.1103/PhysRevResearch.2.013088} {\bibfield  {journal}
  {\bibinfo  {journal} {Phys. Rev. Res.}\ }\textbf {\bibinfo {volume} {2}},\
  \bibinfo {pages} {013088} (\bibinfo {year} {2020})}\BibitemShut {NoStop}%
\bibitem [{\citenamefont {Newnham}(2005)}]{newnham_symmetry}%
  \BibitemOpen
  \bibfield  {author} {\bibinfo {author} {\bibfnamefont {R.~E.}\ \bibnamefont
  {Newnham}},\ }\href@noop {} {\emph {\bibinfo {title} {Properties of
  materials: anisotropy, symmetry, structure}}}\ (\bibinfo  {publisher} {Oxford
  university press},\ \bibinfo {year} {2005})\BibitemShut {NoStop}%
\bibitem [{\citenamefont {Gallego}\ \emph {et~al.}(2019)\citenamefont
  {Gallego}, \citenamefont {Etxebarria}, \citenamefont {Elcoro}, \citenamefont
  {Tasci},\ and\ \citenamefont {Perez-Mato}}]{Gallego_cryst19}%
  \BibitemOpen
  \bibfield  {author} {\bibinfo {author} {\bibfnamefont {S.~V.}\ \bibnamefont
  {Gallego}}, \bibinfo {author} {\bibfnamefont {J.}~\bibnamefont {Etxebarria}},
  \bibinfo {author} {\bibfnamefont {L.}~\bibnamefont {Elcoro}}, \bibinfo
  {author} {\bibfnamefont {E.~S.}\ \bibnamefont {Tasci}},\ and\ \bibinfo
  {author} {\bibfnamefont {J.~M.}\ \bibnamefont {Perez-Mato}},\ }\bibfield
  {title} {\bibinfo {title} {{Automatic calculation of symmetry-adapted tensors
  in magnetic and non-magnetic materials: a new tool of the Bilbao
  Crystallographic Server}},\ }\href
  {https://doi.org/10.1107/S2053273319001748} {\bibfield  {journal} {\bibinfo
  {journal} {Acta Crystallographica Section A}\ }\textbf {\bibinfo {volume}
  {75}},\ \bibinfo {pages} {438} (\bibinfo {year} {2019})}\BibitemShut
  {NoStop}%
\bibitem [{Note2()}]{Note2}%
  \BibitemOpen
  \bibinfo {note} {We emphasize that these quantities are not ${\protect \cal
  C}_{3z}$ symmetric even without any in-plane strain~\cite {Jin_prbL21,
  Jiang_prb20}. However, they are valley contrasting due to the global
  $\protect \cal T$-symmetry of BLG.}\BibitemShut {Stop}%
\bibitem [{Note3()}]{Note3}%
  \BibitemOpen
  \bibinfo {note} {This is a consequence of the strain-induced ${\protect \cal
  M}_x$ symmetry breaking, which makes $\chi _{xx;xy}, \protect \,\chi
  _{yx;xx}$ and $\chi _{yx;yy}$ components finite, modifying the angular
  dependence following Eqs.~\protect \textup {\hbox {\mathsurround \z@ \protect
  \normalfont (\ignorespaces \ref {sigma_long}\unskip \@@italiccorr )}} and
  \protect \textup {\hbox {\mathsurround \z@ \protect \normalfont
  (\ignorespaces \ref {sigma_hall}\unskip \@@italiccorr )}}.}\BibitemShut
  {Stop}%
\bibitem [{\citenamefont {Zheng}\ \emph {et~al.}(2024)\citenamefont {Zheng},
  \citenamefont {Zhai}, \citenamefont {Xiao},\ and\ \citenamefont
  {Yao}}]{Wang_24layer}%
  \BibitemOpen
  \bibfield  {author} {\bibinfo {author} {\bibfnamefont {H.}~\bibnamefont
  {Zheng}}, \bibinfo {author} {\bibfnamefont {D.}~\bibnamefont {Zhai}},
  \bibinfo {author} {\bibfnamefont {C.}~\bibnamefont {Xiao}},\ and\ \bibinfo
  {author} {\bibfnamefont {W.}~\bibnamefont {Yao}},\ }\href@noop {} {\bibinfo
  {title} {Layer coherence origin of intrinsic planar hall effect in 2d limit}}
  (\bibinfo {year} {2024}),\ \Eprint {https://arxiv.org/abs/2402.17166}
  {arXiv:2402.17166 [cond-mat.mes-hall]} \BibitemShut {NoStop}%
\bibitem [{\citenamefont {Taskin}\ \emph {et~al.}(2017)\citenamefont {Taskin},
  \citenamefont {Legg}, \citenamefont {Yang}, \citenamefont {Sasaki},
  \citenamefont {Kanai}, \citenamefont {Matsumoto}, \citenamefont {Rosch},\
  and\ \citenamefont {Ando}}]{Taskin_NC17}%
  \BibitemOpen
  \bibfield  {author} {\bibinfo {author} {\bibfnamefont {A.~A.}\ \bibnamefont
  {Taskin}}, \bibinfo {author} {\bibfnamefont {H.~F.}\ \bibnamefont {Legg}},
  \bibinfo {author} {\bibfnamefont {F.}~\bibnamefont {Yang}}, \bibinfo {author}
  {\bibfnamefont {S.}~\bibnamefont {Sasaki}}, \bibinfo {author} {\bibfnamefont
  {Y.}~\bibnamefont {Kanai}}, \bibinfo {author} {\bibfnamefont
  {K.}~\bibnamefont {Matsumoto}}, \bibinfo {author} {\bibfnamefont
  {A.}~\bibnamefont {Rosch}},\ and\ \bibinfo {author} {\bibfnamefont
  {Y.}~\bibnamefont {Ando}},\ }\bibfield  {title} {\bibinfo {title} {Planar
  hall effect from the surface of topological insulators},\ }\href
  {https://doi.org/10.1038/s41467-017-01474-8} {\bibfield  {journal} {\bibinfo
  {journal} {Nature Communications}\ }\textbf {\bibinfo {volume} {8}},\
  \bibinfo {pages} {1340} (\bibinfo {year} {2017})}\BibitemShut {NoStop}%
\bibitem [{\citenamefont {Breunig}\ \emph {et~al.}(2017)\citenamefont
  {Breunig}, \citenamefont {Wang}, \citenamefont {Taskin}, \citenamefont {Lux},
  \citenamefont {Rosch},\ and\ \citenamefont {Ando}}]{ando_nc17}%
  \BibitemOpen
  \bibfield  {author} {\bibinfo {author} {\bibfnamefont {O.}~\bibnamefont
  {Breunig}}, \bibinfo {author} {\bibfnamefont {Z.}~\bibnamefont {Wang}},
  \bibinfo {author} {\bibfnamefont {A.~A.}\ \bibnamefont {Taskin}}, \bibinfo
  {author} {\bibfnamefont {J.}~\bibnamefont {Lux}}, \bibinfo {author}
  {\bibfnamefont {A.}~\bibnamefont {Rosch}},\ and\ \bibinfo {author}
  {\bibfnamefont {Y.}~\bibnamefont {Ando}},\ }\bibfield  {title} {\bibinfo
  {title} {Gigantic negative magnetoresistance in the bulk of a disordered
  topological insulator},\ }\href {https://doi.org/10.1038/ncomms15545}
  {\bibfield  {journal} {\bibinfo  {journal} {Nature Communications}\ }\textbf
  {\bibinfo {volume} {8}},\ \bibinfo {pages} {15545} (\bibinfo {year}
  {2017})}\BibitemShut {NoStop}%
\bibitem [{\citenamefont {He}\ \emph {et~al.}(2018)\citenamefont {He},
  \citenamefont {Zhang}, \citenamefont {Zhu}, \citenamefont {Liu},
  \citenamefont {Wang}, \citenamefont {Yu}, \citenamefont {Vignale},\ and\
  \citenamefont {Yang}}]{He_NP18}%
  \BibitemOpen
  \bibfield  {author} {\bibinfo {author} {\bibfnamefont {P.}~\bibnamefont
  {He}}, \bibinfo {author} {\bibfnamefont {S.~S.-L.}\ \bibnamefont {Zhang}},
  \bibinfo {author} {\bibfnamefont {D.}~\bibnamefont {Zhu}}, \bibinfo {author}
  {\bibfnamefont {Y.}~\bibnamefont {Liu}}, \bibinfo {author} {\bibfnamefont
  {Y.}~\bibnamefont {Wang}}, \bibinfo {author} {\bibfnamefont {J.}~\bibnamefont
  {Yu}}, \bibinfo {author} {\bibfnamefont {G.}~\bibnamefont {Vignale}},\ and\
  \bibinfo {author} {\bibfnamefont {H.}~\bibnamefont {Yang}},\ }\bibfield
  {title} {\bibinfo {title} {Bilinear magnetoelectric resistance as a probe of
  three-dimensional spin texture in topological surface states},\ }\href
  {https://doi.org/10.1038/s41567-017-0039-y} {\bibfield  {journal} {\bibinfo
  {journal} {Nature Physics}\ }\textbf {\bibinfo {volume} {14}},\ \bibinfo
  {pages} {495–499} (\bibinfo {year} {2018})}\BibitemShut {NoStop}%
\bibitem [{\citenamefont {Wu}\ \emph {et~al.}(2018)\citenamefont {Wu},
  \citenamefont {Pan}, \citenamefont {Wu}, \citenamefont {Fei}, \citenamefont
  {Chen}, \citenamefont {Liu}, \citenamefont {Bu}, \citenamefont {Cao},
  \citenamefont {Song},\ and\ \citenamefont {Wang}}]{Wu_apl18}%
  \BibitemOpen
  \bibfield  {author} {\bibinfo {author} {\bibfnamefont {B.}~\bibnamefont
  {Wu}}, \bibinfo {author} {\bibfnamefont {X.-C.}\ \bibnamefont {Pan}},
  \bibinfo {author} {\bibfnamefont {W.}~\bibnamefont {Wu}}, \bibinfo {author}
  {\bibfnamefont {F.}~\bibnamefont {Fei}}, \bibinfo {author} {\bibfnamefont
  {B.}~\bibnamefont {Chen}}, \bibinfo {author} {\bibfnamefont {Q.}~\bibnamefont
  {Liu}}, \bibinfo {author} {\bibfnamefont {H.}~\bibnamefont {Bu}}, \bibinfo
  {author} {\bibfnamefont {L.}~\bibnamefont {Cao}}, \bibinfo {author}
  {\bibfnamefont {F.}~\bibnamefont {Song}},\ and\ \bibinfo {author}
  {\bibfnamefont {B.}~\bibnamefont {Wang}},\ }\bibfield  {title} {\bibinfo
  {title} {{Oscillating planar Hall response in bulk crystal of topological
  insulator Sn doped Bi1.1Sb0.9Te2S}},\ }\href
  {https://doi.org/10.1063/1.5031906} {\bibfield  {journal} {\bibinfo
  {journal} {Applied Physics Letters}\ }\textbf {\bibinfo {volume} {113}},\
  \bibinfo {pages} {011902} (\bibinfo {year} {2018})}\BibitemShut {NoStop}%
\bibitem [{\citenamefont {He}\ \emph {et~al.}(2019)\citenamefont {He},
  \citenamefont {Zhang}, \citenamefont {Zhu}, \citenamefont {Shi},
  \citenamefont {Heinonen}, \citenamefont {Vignale},\ and\ \citenamefont
  {Yang}}]{He_prl19}%
  \BibitemOpen
  \bibfield  {author} {\bibinfo {author} {\bibfnamefont {P.}~\bibnamefont
  {He}}, \bibinfo {author} {\bibfnamefont {S.~S.-L.}\ \bibnamefont {Zhang}},
  \bibinfo {author} {\bibfnamefont {D.}~\bibnamefont {Zhu}}, \bibinfo {author}
  {\bibfnamefont {S.}~\bibnamefont {Shi}}, \bibinfo {author} {\bibfnamefont
  {O.~G.}\ \bibnamefont {Heinonen}}, \bibinfo {author} {\bibfnamefont
  {G.}~\bibnamefont {Vignale}},\ and\ \bibinfo {author} {\bibfnamefont
  {H.}~\bibnamefont {Yang}},\ }\bibfield  {title} {\bibinfo {title} {Nonlinear
  planar hall effect},\ }\href {https://doi.org/10.1103/PhysRevLett.123.016801}
  {\bibfield  {journal} {\bibinfo  {journal} {Phys. Rev. Lett.}\ }\textbf
  {\bibinfo {volume} {123}},\ \bibinfo {pages} {016801} (\bibinfo {year}
  {2019})}\BibitemShut {NoStop}%
\bibitem [{\citenamefont {Zheng}\ \emph
  {et~al.}(2020{\natexlab{a}})\citenamefont {Zheng}, \citenamefont {Duan},
  \citenamefont {Wang}, \citenamefont {Li}, \citenamefont {Deng},\ and\
  \citenamefont {Wang}}]{Wang_prbR20}%
  \BibitemOpen
  \bibfield  {author} {\bibinfo {author} {\bibfnamefont {S.-H.}\ \bibnamefont
  {Zheng}}, \bibinfo {author} {\bibfnamefont {H.-J.}\ \bibnamefont {Duan}},
  \bibinfo {author} {\bibfnamefont {J.-K.}\ \bibnamefont {Wang}}, \bibinfo
  {author} {\bibfnamefont {J.-Y.}\ \bibnamefont {Li}}, \bibinfo {author}
  {\bibfnamefont {M.-X.}\ \bibnamefont {Deng}},\ and\ \bibinfo {author}
  {\bibfnamefont {R.-Q.}\ \bibnamefont {Wang}},\ }\bibfield  {title} {\bibinfo
  {title} {Origin of planar hall effect on the surface of topological
  insulators: Tilt of dirac cone by an in-plane magnetic field},\ }\href
  {https://doi.org/10.1103/PhysRevB.101.041408} {\bibfield  {journal} {\bibinfo
   {journal} {Phys. Rev. B}\ }\textbf {\bibinfo {volume} {101}},\ \bibinfo
  {pages} {041408} (\bibinfo {year} {2020}{\natexlab{a}})}\BibitemShut
  {NoStop}%
\bibitem [{\citenamefont {Rao}\ \emph {et~al.}(2021)\citenamefont {Rao},
  \citenamefont {Zhou}, \citenamefont {Wu}, \citenamefont {Duan}, \citenamefont
  {Deng},\ and\ \citenamefont {Wang}}]{Wang_prb21}%
  \BibitemOpen
  \bibfield  {author} {\bibinfo {author} {\bibfnamefont {W.}~\bibnamefont
  {Rao}}, \bibinfo {author} {\bibfnamefont {Y.-L.}\ \bibnamefont {Zhou}},
  \bibinfo {author} {\bibfnamefont {Y.-j.}\ \bibnamefont {Wu}}, \bibinfo
  {author} {\bibfnamefont {H.-J.}\ \bibnamefont {Duan}}, \bibinfo {author}
  {\bibfnamefont {M.-X.}\ \bibnamefont {Deng}},\ and\ \bibinfo {author}
  {\bibfnamefont {R.-Q.}\ \bibnamefont {Wang}},\ }\bibfield  {title} {\bibinfo
  {title} {Theory for linear and nonlinear planar hall effect in topological
  insulator thin films},\ }\href {https://doi.org/10.1103/PhysRevB.103.155415}
  {\bibfield  {journal} {\bibinfo  {journal} {Phys. Rev. B}\ }\textbf {\bibinfo
  {volume} {103}},\ \bibinfo {pages} {155415} (\bibinfo {year}
  {2021})}\BibitemShut {NoStop}%
\bibitem [{\citenamefont {Ai}\ \emph {et~al.}(2024)\citenamefont {Ai},
  \citenamefont {Chen}, \citenamefont {Liu}, \citenamefont {Yuan},
  \citenamefont {Zhang}, \citenamefont {He}, \citenamefont {Dong},
  \citenamefont {Fu}, \citenamefont {Luo}, \citenamefont {Deng}, \citenamefont
  {Wang},\ and\ \citenamefont {Wu}}]{Jinxiong_NC24}%
  \BibitemOpen
  \bibfield  {author} {\bibinfo {author} {\bibfnamefont {W.}~\bibnamefont
  {Ai}}, \bibinfo {author} {\bibfnamefont {F.}~\bibnamefont {Chen}}, \bibinfo
  {author} {\bibfnamefont {Z.}~\bibnamefont {Liu}}, \bibinfo {author}
  {\bibfnamefont {X.}~\bibnamefont {Yuan}}, \bibinfo {author} {\bibfnamefont
  {L.}~\bibnamefont {Zhang}}, \bibinfo {author} {\bibfnamefont
  {Y.}~\bibnamefont {He}}, \bibinfo {author} {\bibfnamefont {X.}~\bibnamefont
  {Dong}}, \bibinfo {author} {\bibfnamefont {H.}~\bibnamefont {Fu}}, \bibinfo
  {author} {\bibfnamefont {F.}~\bibnamefont {Luo}}, \bibinfo {author}
  {\bibfnamefont {M.}~\bibnamefont {Deng}}, \bibinfo {author} {\bibfnamefont
  {R.}~\bibnamefont {Wang}},\ and\ \bibinfo {author} {\bibfnamefont
  {J.}~\bibnamefont {Wu}},\ }\bibfield  {title} {\bibinfo {title} {Observation
  of giant room-temperature anisotropic magnetoresistance in the topological
  insulator $\beta$-ag2te},\ }\href
  {https://doi.org/10.1038/s41467-024-45643-y} {\bibfield  {journal} {\bibinfo
  {journal} {Nature Communications}\ }\textbf {\bibinfo {volume} {15}},\
  \bibinfo {pages} {1259} (\bibinfo {year} {2024})}\BibitemShut {NoStop}%
\bibitem [{\citenamefont {Zheng}\ \emph
  {et~al.}(2020{\natexlab{b}})\citenamefont {Zheng}, \citenamefont {Ma},
  \citenamefont {Bi}, \citenamefont {de~la Barrera}, \citenamefont {Liu},
  \citenamefont {Mao}, \citenamefont {Zhang}, \citenamefont {Kiper},
  \citenamefont {Watanabe}, \citenamefont {Taniguchi}, \citenamefont {Kong},
  \citenamefont {Tisdale}, \citenamefont {Ashoori}, \citenamefont {Gedik},
  \citenamefont {Fu}, \citenamefont {Xu},\ and\ \citenamefont
  {Jarillo-Herrero}}]{Zheng_Nat20}%
  \BibitemOpen
  \bibfield  {author} {\bibinfo {author} {\bibfnamefont {Z.}~\bibnamefont
  {Zheng}}, \bibinfo {author} {\bibfnamefont {Q.}~\bibnamefont {Ma}}, \bibinfo
  {author} {\bibfnamefont {Z.}~\bibnamefont {Bi}}, \bibinfo {author}
  {\bibfnamefont {S.}~\bibnamefont {de~la Barrera}}, \bibinfo {author}
  {\bibfnamefont {M.-H.}\ \bibnamefont {Liu}}, \bibinfo {author} {\bibfnamefont
  {N.}~\bibnamefont {Mao}}, \bibinfo {author} {\bibfnamefont {Y.}~\bibnamefont
  {Zhang}}, \bibinfo {author} {\bibfnamefont {N.}~\bibnamefont {Kiper}},
  \bibinfo {author} {\bibfnamefont {K.}~\bibnamefont {Watanabe}}, \bibinfo
  {author} {\bibfnamefont {T.}~\bibnamefont {Taniguchi}}, \bibinfo {author}
  {\bibfnamefont {J.}~\bibnamefont {Kong}}, \bibinfo {author} {\bibfnamefont
  {W.~A.}\ \bibnamefont {Tisdale}}, \bibinfo {author} {\bibfnamefont
  {R.}~\bibnamefont {Ashoori}}, \bibinfo {author} {\bibfnamefont
  {N.}~\bibnamefont {Gedik}}, \bibinfo {author} {\bibfnamefont
  {L.}~\bibnamefont {Fu}}, \bibinfo {author} {\bibfnamefont {S.-Y.}\
  \bibnamefont {Xu}},\ and\ \bibinfo {author} {\bibfnamefont {P.}~\bibnamefont
  {Jarillo-Herrero}},\ }\bibfield  {title} {\bibinfo {title} {Unconventional
  ferroelectricity in moir{\'e} heterostructures},\ }\href
  {https://doi.org/10.1038/s41586-020-2970-9} {\bibfield  {journal} {\bibinfo
  {journal} {Nature}\ }\textbf {\bibinfo {volume} {588}},\ \bibinfo {pages}
  {71} (\bibinfo {year} {2020}{\natexlab{b}})}\BibitemShut {NoStop}%
\bibitem [{\citenamefont {Wang}\ \emph {et~al.}(2022)\citenamefont {Wang},
  \citenamefont {Yasuda}, \citenamefont {Zhang}, \citenamefont {Liu},
  \citenamefont {Watanabe}, \citenamefont {Taniguchi}, \citenamefont {Hone},
  \citenamefont {Fu},\ and\ \citenamefont {Jarillo-Herrero}}]{Wang_NatNano22}%
  \BibitemOpen
  \bibfield  {author} {\bibinfo {author} {\bibfnamefont {X.}~\bibnamefont
  {Wang}}, \bibinfo {author} {\bibfnamefont {K.}~\bibnamefont {Yasuda}},
  \bibinfo {author} {\bibfnamefont {Y.}~\bibnamefont {Zhang}}, \bibinfo
  {author} {\bibfnamefont {S.}~\bibnamefont {Liu}}, \bibinfo {author}
  {\bibfnamefont {K.}~\bibnamefont {Watanabe}}, \bibinfo {author}
  {\bibfnamefont {T.}~\bibnamefont {Taniguchi}}, \bibinfo {author}
  {\bibfnamefont {J.}~\bibnamefont {Hone}}, \bibinfo {author} {\bibfnamefont
  {L.}~\bibnamefont {Fu}},\ and\ \bibinfo {author} {\bibfnamefont
  {P.}~\bibnamefont {Jarillo-Herrero}},\ }\bibfield  {title} {\bibinfo {title}
  {Interfacial ferroelectricity in rhombohedral-stacked bilayer transition
  metal dichalcogenides},\ }\href {https://doi.org/10.1038/s41565-021-01059-z}
  {\bibfield  {journal} {\bibinfo  {journal} {Nature Nanotechnology}\ }\textbf
  {\bibinfo {volume} {17}},\ \bibinfo {pages} {367} (\bibinfo {year}
  {2022})}\BibitemShut {NoStop}%
\bibitem [{\citenamefont {Yasuda}\ \emph {et~al.}(2021)\citenamefont {Yasuda},
  \citenamefont {Wang}, \citenamefont {Watanabe}, \citenamefont {Taniguchi},\
  and\ \citenamefont {Jarillo-Herrero}}]{Herrero_sci21}%
  \BibitemOpen
  \bibfield  {author} {\bibinfo {author} {\bibfnamefont {K.}~\bibnamefont
  {Yasuda}}, \bibinfo {author} {\bibfnamefont {X.}~\bibnamefont {Wang}},
  \bibinfo {author} {\bibfnamefont {K.}~\bibnamefont {Watanabe}}, \bibinfo
  {author} {\bibfnamefont {T.}~\bibnamefont {Taniguchi}},\ and\ \bibinfo
  {author} {\bibfnamefont {P.}~\bibnamefont {Jarillo-Herrero}},\ }\bibfield
  {title} {\bibinfo {title} {Stacking-engineered ferroelectricity in bilayer
  boron nitride},\ }\href {https://doi.org/10.1126/science.abd3230} {\bibfield
  {journal} {\bibinfo  {journal} {Science}\ }\textbf {\bibinfo {volume}
  {372}},\ \bibinfo {pages} {1458} (\bibinfo {year} {2021})}\BibitemShut
  {NoStop}%
\end{thebibliography}%
\end{document}